\documentclass[aps,notitlepage,pre,showpacs,showkeys,amsmath,amsfonts,amssymb,nofootinbib]{revtex4-1}

\usepackage{graphicx}
\usepackage{subfigure}
\usepackage[titletoc,toc,title,page]{appendix}
\usepackage{placeins}
\usepackage{color}
\usepackage{hyperref}
\usepackage{dcolumn}
\usepackage{bm}

\newcommand{\be}{\begin{equation}}
\newcommand{\ee}{\end{equation}}
\newcommand{\ba}{\begin{eqnarray}}
\newcommand{\ea}{\end{eqnarray}}
\newcommand{\s}{\sigma}
\newcommand{\la}{\lambda}
\newcommand{\fr}{\frac{1}{2}}
\newcommand{\p}{\partial}
\newcommand{\frs}{\frac{2}{\s^2}}
\newcommand{\frsi}{\frac{\s^2}{2}}
\newcommand{\frii}{\frac{\sqrt{2}}{\s}}
\newcommand{\Aa}{\mathcal A}
\newcommand{\Bb}{\mathcal B}

\newcommand{\Lim}[1]{\raisebox{0.5ex}{\scalebox{0.8}{$\displaystyle \lim_{#1}\;$}}}

\begin{document}

\preprint{preprint}

\title[F. Guarnieri]{Solution of the Fokker-Planck equation with a logarithmic potential and mixed eigenvalue spectrum\\}

\author{F. Guarnieri}
\affiliation{Nordic Institute for Theoretical Physics (NORDITA), 106 91 Stockholm, Sweden}
\email{filippo.guarnieri@roma1.infn.it}

\author{W. Moon}
\affiliation{British Antarctic Survey, High Cross, Madingley Road,
Cambridge, CB3 0ET, UK}%
%
\email{woosok.moon@gmail.com}

\author{J. S. Wettlaufer}%
\affiliation{ 
Yale University, New Haven, CT, 06520-8109, USA
}%
\affiliation{Mathematical Institute, University of Oxford, Oxford
OX2 6GG, UK}
\affiliation{Nordic Institute for Theoretical Physics (NORDITA), 106 91 Stockholm, Sweden
}%
\email{john.wettlaufer@yale.edu}

             
\begin{abstract}
Motivated by a problem in climate dynamics, we investigate the solution of a Bessel-like process with negative constant drift, described by a Fokker-Planck equation with a potential $V(x) = - [b \ln(x) + a\, x]$, for $b>0$ and $a<0$. The problem belongs to a family of Fokker-Planck equations with logarithmic potentials closely related to the Bessel process, that has been extensively studied for its applications in physics, biology and finance. The {\em Bessel-like} process we consider can be solved by seeking solutions through an expansion into a complete set of eigenfunctions. The associated imaginary-time Schr\"odinger equation exhibits a mix of discrete and continuous eigenvalue spectra, corresponding to the quantum Coulomb potential describing the bound states of the hydrogen atom. We present a technique to evaluate the normalization factor of the continuous spectrum of eigenfunctions that relies solely upon their asymptotic behavior. We demonstrate the technique by solving the Brownian motion problem and the Bessel process both with a negative constant drift. We conclude with a comparison with other analytical methods and with numerical solutions. 
\end{abstract}


\maketitle

\vspace{-0.2in}
\section{\label{sec:intro}Introduction}
Many complex systems can be investigated by capturing the effect of a large number of degrees of freedom by adding a noise term to a deterministic evolution equation. This may be done either phenomenologically or by implementing a stochastic reduction procedure on a more complex set of equations \cite[e.g.][]{franzke2005, MW2017}.  Statistical information is governed by the probability density function $\rho(x,t)$, which satisfies the Fokker-Planck equation
\be\label{FPE}
\p_t\, \rho(x,t) = - \, \p_x\, \left[ f(x)\, \rho(x,t) \right] + \fr\, \s^2\, \p_x^2\, \rho(x,t) \, ,
\ee
a second order parabolic partial differential equation wherein $f(x)$ describes the deterministic dynamics and $\s$ is the constant noise strength. (Note, for parsimony of notation, we write the operator $\p^2/\p {x^2}$ as $\p_x^2$.) For example, when a one-dimensional solely diffusive process governs $\rho(x,t)$, there is no deterministic drift force ($f(x) = 0$), and the Fokker-Planck equation is a diffusion equation with a Gaussian solution; 
\be\label{Brownian}
\rho(x,t) = \frac{1}{(2\, \pi\, \s^2\, t)^\frac{1}{2}}\, \text{e}^{- \frac{x^2}{2\, \s^2\, t}}\, ,
\ee
for initial data $\rho(x,t=0)=\delta(x)$, with natural boundary conditions $\rho(x=0,t)=\rho(x=\infty,t)=0$\footnote{Here natural, absorbing and reflecting boundaries refer to boundaries where the probability density vanishes (sufficiently fast to insure normalization), vanishes with finite flux, or has zero flux, respectively.  Typically, natural boundary conditions are imposed at $\pm \infty$.  See also Feller \cite{FELL1952}}.
The long-time behavior of Eq. \eqref{FPE} for $f(x) \neq 0$ is given by the Boltzmann distribution in the case of thermal systems, defined as 
\be\label{SS}
\rho_{st}(x) = \frac{\mathcal N}{\s^2}\, \text{e}^{-\frac{2}{\s^2}\,V(x)} \, ,
\ee
where $\mathcal N$ is a normalization constant and we can consider $V(x) = - \int^x f(y)\, dy$ to be a potential in which, say, a Brownian particle evolves and hence the connotation of the deterministic drift force; $f(x) = - d_x V(x)$. 

A large class of systems are described by a logarithmic potential (sometimes referred to as an entropy term), viz., $V(x)= b\,\ln(x)$, with $b$ constant. Depending on the sign of $b$, the central force associated with this potential is either attractive or repulsive, and systems governed by this potential are generally referred to as Bessel processes, characterized by the Fokker-Planck equation
\be\label{FPEB}
\p_t\, \rho(x,t) = - \, \p_x\, \left[- \frac{b}{x}\, \rho(x,t)\right]  + \fr\, \s^2\, \p_x^2\, \rho(x,t) \, ,
\ee
which has the following exact solution 
\be\label{EXBess}
\rho(x,t) = x^{\left(\fr-\frac{b}{\s^2}\right)}\, x_0^{\fr + \frac{b}{\s^2}}\, \frac{1}{\s^2\, t}\, \text{e}^{- \frac{x^2+x_0^2}{2\, \s^2\, t}} \, I_{\fr + \frac{b}{\s^2}}\left(\frac{x\,x_0}{\s^2\, t}\right) \, ,
\ee
in which $I_{\fr + \frac{b}{\s^2}}$ is the modified Bessel function of the first kind, and $x_0 = x(t=0)$.  
Due to the singularity at the origin, Eq.\eqref{FPEB} is defined on the half-line [0,$\infty$), the origin satisfies a reflecting boundary condition and there is a natural boundary condition $\rho(x=\infty,t) = 0$ at infinity. 

The Bessel process exhibits several interesting properties. For example, as described in \cite[refs.][]{BRWL2000,MBGB2010}, by integrating over angle variables and letting $b \to b - (d-1)\,\s^2/2$, the equation for the radial component in $d$-dimensions is equivalent to the one-dimensional form given by Eq. \eqref{FPEB}. Thus, a particle in a logarithmic potential in a space dimension $d$ will have the same radial distribution function as a free-particle in $\tilde d = d -2\, b/\s^2$ dimensions, where $\tilde d$ can be negative and non-integer.  Pitman and Yor \cite{PIYO1981,PIYO1982} studied the relationships between Brownian motion and Bessel processes, and squared Bessel processes are connected to local times (or occupation times) in Brownian motion through a space-time transformations and the application of Ray-Knight theorems \cite[e.g.,][]{YORE1991,YOMA1999}.   Hence,  squared Bessel processes play a major role in the study of maps between Brownian motion in different dimensions \cite{WILL1970,PITM1975}. 

The importance of Bessel diffusion processes  is generally two-fold.  Firstly, it belongs to a restricted class of exactly solvable models, along with Brownian motion and the Ornstein-Uhlenbeck process. Secondly, it is widely applicable in physics, chemistry, biology and finance. For example, in physics, it describes (a) the effective long-range interaction between two probe particles in a one-dimensional driven fluid, such as solute particles in a solvent \cite{LMSL2005}; (b) the vortex-antivortex annihilation in the 2D XY Heisenberg model subject to thermal fluctuations at temperatures below the Kosterlitz-Thouless transition $T_{KT}$ \cite{BRWL2000}; (c) an alternative technique to quantize non-Abelian gauge field theories, avoiding the drawbacks of the Fadeev-Popov method \cite{HHSA1983}\footnote{Note that in Ref. \cite[][]{HHSA1983} the Bessel process is mistakenly referred to as the Rayleigh process.}. In finance, the property that the Bessel process is equivalent to exponential Brownian motion with linear drift in time (or geometric Brownian motion) makes it a suitable tool for the evaluation of Asian options, which are contracts based on the average underlying price and are thus considered less expensive than standard options, due to the lower variance of the averages \cite{GEYO1993,GJYO2003}. 

As summarized nicely by Dechant \emph{et al.}, \citep{DLBK2011}, because the Bessel process with a purely logarithmic potential $V(x)= b\,\ln|x|$ is singular at the origin, then the steady state given by the Boltzmann distribution of Eq. (\ref{SS}) is non-integrable.  This is because $\rho_{st}(x) \propto \text{e}^{-\frac{2}{\s^2}\,V(x)} = x^{- 2\, b/\s^2}$ clearly diverges at large $x$ (small $x$) depending on whether $2\, b/\s^2 < 1$ ($2\, b/\s^2 > 1$).  However, it is more common that the potential of real systems only behaves  like a pure logarithm asymptotically. Therefore, it is more appropriate to talk about ``Bessel-like processes''. For example, in the settings of (a) and (b) discussed above, there is an effective characteristic scale $a$ such that at large $x$, the potential behaves as $V(x) \approx b\, \ln(x/a)$. In other settings, such as in the case of the diffusive spreading of momenta in two-level atoms in optical lattices \cite{DGRT2006}, the potential may also exhibit a more complex logarithmic dependence, such as
\be
V(x) = \frac{b}{2}\, \ln(1 + x^2)\, ,\qquad \text{and hence} \qquad f(x) = - \frac{b\, x}{1+x^2}\, ,
\ee
where $x$ is the momentum. 

Because of the slow power law spatial decay, few moments of the stationary distribution will diverge. Therefore, complete information regarding the stationary distribution is not provided by the time-independent solution $\rho_{st}$, but instead requires taking the infinite-time limit of the time-dependent moments \cite{DLBK2011}. The latter has been proved by Lutz to be related to a breaking in the ergodicity for systems with power-law distributions \cite{LUTP2004}.  
By adding regular terms to the potential, the
existence of the stationary solution can always be guaranteed.  One example is the Rayleigh process \cite{RAYL1880,RAYL1905,SNTR1963,GNRS1986}, given by
\be\label{Rayl}
V(x) = b\, \ln(x) + \frac{a}{2}\, x^2 \,, \qquad \text{and hence} \qquad f(x) = - \frac{b}{x} - a\, x \,,
\ee
which is considered to be a generalized Bessel process, or a Bessel process with a linear drift or radial Ornstein-Uhlenbeck process, as it generalizes the radial projection of a $d$-dimensional Ornstein-Uhlenbeck equation \cite{ORUH1930}. The Rayleigh process is widely used in finance because the parabolic term corresponds to the Cox-Ingersoll-Ross (CIR) process \cite{CIRO1985}. The CIR process describes the evolution of financial variables that are strictly positive and empirically known to exhibit mean reversion, in which there is a convergence to the average value in the long time limit \cite{GJYO2003}. 

Here we focus on another generalization of the Bessel process wherein the drift is characterized by the potential $V(x) = - b\ln(x) - a\, x$, and is thus associated with the Fokker-Planck equation,
\be\label{FPEMINE}
\p_t\, \rho(x,t) = - \p_x\, \left[\left(\frac{b}{x} + a \right)\, \rho(x,t)\right]  + \fr\, \s^2\, \p_x^2\, \rho(x,t) \, ,
\ee
which arises in applications in biology, queuing theory, finance and climate dynamics. In biology, it models the dynamics of DNA bubbles forming when the hydrogen bonds of the base pairs in the Watson-Crick double-helix structure break due to thermal fluctuations \cite{FMDD2007}. The continuous zipping and unzipping of the double-strand, referred to as ``DNA breathing'', allows chemicals and proteins to bind to reactive sites of the bases that they would not otherwise be able to access \cite{FMDD2007}. Thus, when $x$ in Eq.\eqref{FPEMINE} refers to the dynamics of the bubble length, this is referred to as the Poland-Scheraga model. In queueing theory, Cofmann \emph{et al.} \citep{CPRE1998} contrast the relevance of the Bessel process and reflective Brownian motion--a random walk with a reflective singularity at the origin.  The former (latter) case describes the distribution of the unfinished work in the heavy-traffic limit for a two queue system served by a single server that switches between them instantaneously (over a finite time).  
In finance, this process has been applied to the evaluation of Asian options \cite{VALI2004}. Linetsky \cite{LINE2004} uses a non-linear transformation and a measure change of Eq. \eqref{FPEMINE} to find a generalization of the CIR process that provides a better fit of empirical data.  In climate dynamics, Toppaladoddi and Wettlaufer have derived a Fokker-Planck-like equation for the distribution of Arctic sea ice thickness that, in geophysically relevant limits, can be described by a Bessel process with constant drift, and have found both stationary and time-dependent solutions to match satellite data \cite{STJW2015, STJW2017}.  The concept of the sea ice thickness distribution was introduced by Thorndike \emph{et al.} \cite{Thorndike:1975}, and is defined by considering a region with area $R$ that is sufficiently large to contain a range ice of different thicknesses. Then the integral
\be
\int_{x_1}^{x_2} \rho(x,t) \, dx = \frac{A}{R}
\ee
gives the fraction of that area ($A/R$) that contains ice of thicknesses between $x_1$ and $x_2$.  Then, for a simple treatment of the thermal growth process, the spatio-temporal evolution of $\rho(x,t)$, subject to wind, thermal and mechanical forcing, is governed by Eq. \eqref{FPEMINE}.   A fully coupled approach to the climatological evolution of the ice thickness distribution uses a thermodynamic only nonlinear nonautonomous deterministic backbone
\cite{EWNT2009} and thus requires a numerical approach.  However, a stochastic perturbation theory has been developed that provides numerical tests of the Langevin equations emerging from a range of generalizations of the geophysical models \cite{MWJMP2013, STJW2015, MWJC2017}.  

Whilst the ``bare'' Bessel process has been solved exactly, the Bessel process with constant drift poses significant challenges in this regard.  Progress can be made by solving the Fokker-Planck equation in terms of either Fourier or Legendre transforms and numerically transforming back \cite{LNLE2014}, or by expanding the solution directly into a complete set of orthonormal eigenfunctions and solving the associated Sturm-Liouville eigenvalue problem. For an infinite domain, the presence of a constant drift term leads to an eigenvalue spectrum that is a mix of a discrete and continuous set of eigenvalues, the latter of which are associated with eigenfunctions that oscillate at infinity, posing obvious complications. Indeed, oscillating solutions must be normalized to a $\delta$-function, and thus to unity in the sense of distributions. The process described by Eq. \eqref{FPEMINE} has been solved with several techniques. Linetsky gave a spectral representation both by solving the equation in real space in a finite box and then sending the box size to infinity \cite{VALI2004}, and by solving the equation in complex space in the full domain $[0,\infty)$ \cite{LINE2004} (see \cite{TITC1962} for an introduction to the technique). We note that the associated Sturm-Liouville equation strongly resembles the Schr\"odinger equation for the Coulomb interaction in quantum mechanics, describing the discrete eigenstates of the hydrogen atom.  Here, motivated by the climate dynamics problem described above, we employ a different technique to solve this problem. The method evaluates the normalization of oscillatory eigenfunctions on the infinite half-line without the use of complex variables, and solely requires knowledge of their asymptotic oscillatory behavior (heuristically like the method of stationary phase, but clearly without appeal to the Riemann-Lebesgue Lemma).
The paper is organized as follows. In \S \ref{sec:FPE} we introduce the spectral expansion of the  Fokker-Planck equation, and describe how to normalize oscillating solutions. In \S \ref{sec:RBM}, we use this method to solve the problem of Brownian motion with constant negative drift.  Finally, in \S \ref{sec:BE} we solve the Bessel process with negative constant negative drift, and compare the results with other analytic techniques and numerical solutions.
\vspace{-0.2in}
\section{\label{sec:FPE}Spectral Expansion of the Fokker-Planck equation}
The general form of the Fokker-Planck equation in the It\^o formulation is
\be\label{FPEnew}
\p_t\, \rho(x,t) = \mathcal{L}\, \rho(x,t)\, ; \qquad \mathcal{L} = - \p_x\, f(x,t) + \fr\, \p_x^2\, g(x,t)^2\,  ,
\ee
where $\mathcal L$ is the Fokker-Planck operator, and $f(x,t)$ and $g(x,t)$ are the drift and diffusion functions respectively.  For given initial data, Eq. \eqref{FPEnew} satisfies conditions at the boundary of the domain $\Omega = [x_{in},x_{out}]$. We will consider the simpler case of additive noise, where $g(x) \equiv \s^2$  defines the constant noise strength, and time-independent constant drift  $f(x,t) \equiv f(x)$, and hence the Fokker-Planck equation becomes
\be\label{FPEnew2}
\p_t\, \rho(x,t) = \mathcal{L}\, \rho(x,t)\, ; \qquad \mathcal{L} = - \p_x\, f(x) + \fr\, \s^2\p_x^2\, \,  ,
\ee
corresponding to which is the autonomous Langevin equation 
\be
dx = f(x)\, dt + \sigma\, dW(t)\, ,
\ee
where $dW(t)$ is a Wiener process. The probability density can be written as
\be\label{probtrans}
\rho(x,t) = \int_\Omega dx_0\, \rho(x,t|x_0,t_0)\, \rho_{in}(x_0,t_0) \, ,
\ee
where $\rho_{in}(x_0,t_0)$ is an initial condition at time $t_0$ and position $x_0$.  The transition density satisfies Eq. \eqref{FPEnew2} with initial condition $\Lim{t\to t_0} \rho(x,t|x_0,t_0) = \delta(x-x_0)$, and thus formally
\be\label{formaltrans}
\rho(x,t|x_0,t_0) = \text{e}^{\mathcal{L}(t-t_0)}\, \delta(x-x_0) \, .
\ee
Thus, as described in \S \ref{sec:intro}, the stationary solution of Eq. \eqref{FPEnew2} is 
\be\label{SS2}
\rho_{st}(x) = \frac{\mathcal N}{\s^2}\, \text{e}^{-\frac{2}{\s^2}\,V(x)} \, ,
\ee
where $\mathcal N$ is a normalization constant, viz., 
\be
\int_\Omega dx\, \rho_{st}(x) = 1 \, ,
\ee
and we can consider $V(x) = - \int^x f(y)\, dy$ to be a potential associated with the deterministic drift force; $f(x) = - d_x V(x)$. 
The existence of a renormalizable stationary solution is guaranteed by requiring that the potential diverges at both boundaries, and hence for example a particle is strictly confined within the domain.
\subsection{\label{sec:EE}Eigenfunction expansion}
The Fokker-Planck equation can be solved exactly just in few cases. Generally, a solution can be obtained in real space by expanding the transition density into a set of eigenfunctions and solving the associated eigenvalue problem \cite{RFPE1934}. We expand the transition density as
\be\label{eig0}
\rho(x,t|x_0,t_0) = \sum_{\la\in\Lambda} \, \text{e}^{- \la\, (t-t_0)}\, \phi_\la(x_0)\, \phi_\la(x) \, ,
\ee
where the $\phi_\la$ are the set of eigenfunctions of $\mathcal L$ satisfying the eigenvalue problem 
\be\label{eigprob}
\mathcal{L}\, \phi_\la(x) = -\la\, \phi_\la(x) \, ,
\ee
and $\Lambda$ is the eigenvalue spectrum. Inserting the expansion \eqref{eig0} into the density \eqref{probtrans} we find
\be\label{eig1}
\rho(x,t) = \int_\Omega dx_0\, \rho(x,t|x_0,t_0) \, \rho_{in}(x_0,t_0)  = \sum_{\la\in\Lambda} \, \alpha_\la(t_0)\, \text{e}^{- \la\, (t-t_0)}\, \phi_\la(x) \, ,
\ee
where the coefficients $\alpha_\la(t_0)$ are the projections of the initial condition on the $\lambda^{\text{th}}$ function, viz.,
\be
\alpha_\la(t_0) = \int_\Omega\, dx_0\, \rho_{in}(x_0,t_0)\, \phi_\la(x_0)  \, .
\ee
The calculation of the spectrum can be simplified by rewriting Eq. \eqref{eigprob} in terms of a self-adjoint operator. The Sturm-Liouville theorem ensures that such operators have a complete basis of orthonormal eigenfunctions and a real spectrum. However, the operator $\mathcal{L}$ in Eq. \eqref{eigprob} is not self-adjoint, which can be seen by writing $\mathcal{L}$ in terms of the potential $V(x)$ as
\be
\mathcal{L} = \frac{\s^2}{2}\, \p_x\, \text{e}^{-\frs V(x)}\, \p_x\, \text{e}^{\frs V(x)} \, 
\ee
and checking that for two test functions $\phi_1$ and $\phi_2$ the following inequality holds
\be
\int_\Omega\, dx\, \phi_1(x)\, \left\{\mathcal{L}\, \phi_2(x)\right\}\,\neq \int_\Omega\, dx\, \phi_2(x)\, \left\{\mathcal{L}\, \phi_1(x)\right\}\, \, .
\ee
Nevertheless, we can construct a self-adjoint operator associated with the Fokker-Planck operator as
\be\label{selfaggio}
\mathcal{\tilde L} = \text{e}^{\frs V(x)}\, \mathcal{L}\, ,
\ee
which satisfies Eq. \eqref{eigprob} and possesses a complete orthonormal set of eigenfunctions $\{\psi_\lambda\}$ with non-negative real eigenvalues $\la \geq 0$. The two operators have the same spectrum, and their eigenfunctions are, up to normalization, related by 
\be\label{repa}
\phi_\la(x) = \text{e}^{-\frac{V(x)}{\s^2}}\, \psi_\la(x) \propto \sqrt{\rho_{st}(x)}\, \psi_\la(x)\,.
\ee
 The completeness of the basis insures that the initial condition of the transition density satisfies
\be\label{complete}
\Lim{t\to t_0}\, \rho(x,t|x_0,t_0) =  \delta(x-x_0) = \sum_{\la\in\Lambda}\, \psi_\la(x_0)\, \psi_\la(x) = \text{e}^{\frac{V(x)}{\s^2}+\frac{V(x_0)}{\s^2}}\, \sum_{\la\in\Lambda}\, \phi_\la(x_0)\, \phi_\la(x) \, ,
\ee
which, by adding unity, in the sense of distributions, can be rewritten as
\ba\label{complete}
\delta(x-x_0) =
\text{e}^{-\frac{V(x)}{\s^2}+\frac{V(x_0)}{\s^2}}\, \delta(x-x_0)\, = \text{e}^{+\frac{V(x)}{\s^2}-\frac{V(x_0)}{\s^2}}\, \delta(x-x_0)\, \\= \text{e}^{\frs V(x)}\, \sum_{\la\in\Lambda}\, \phi_\la(x_0)\, \phi_\la(x)  = \text{e}^{\frs V(x_0)}\, \sum_{\la\in\Lambda}\, \phi_\la(x_0)\, \phi_\la(x) \,.
\ea
Substituting \eqref{complete} into \eqref{formaltrans}, and then into \eqref{eig1} using \eqref{repa}, we find
\be\label{discretefull}
\rho(x,t) = \int_\Omega\, dx_0\, \sum_{\la\in\Lambda}\, \text{e}^{- \la(t-t_0)}\, \sqrt{\frac{\rho_{st}(x)}{\rho_{st}(x_0)}}\, \psi_\la(x_0)\, \psi_\la(x)\, \rho_{in}(x_0,t_0) \, .
\ee
%
\subsection{\label{sec:SE}The Schr\"odinger-like equation}
\begin{center}
\begin{figure}[ht]
\includegraphics[height=4cm]{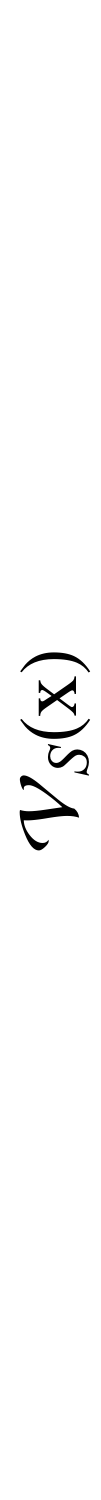}
\includegraphics[height=4cm]{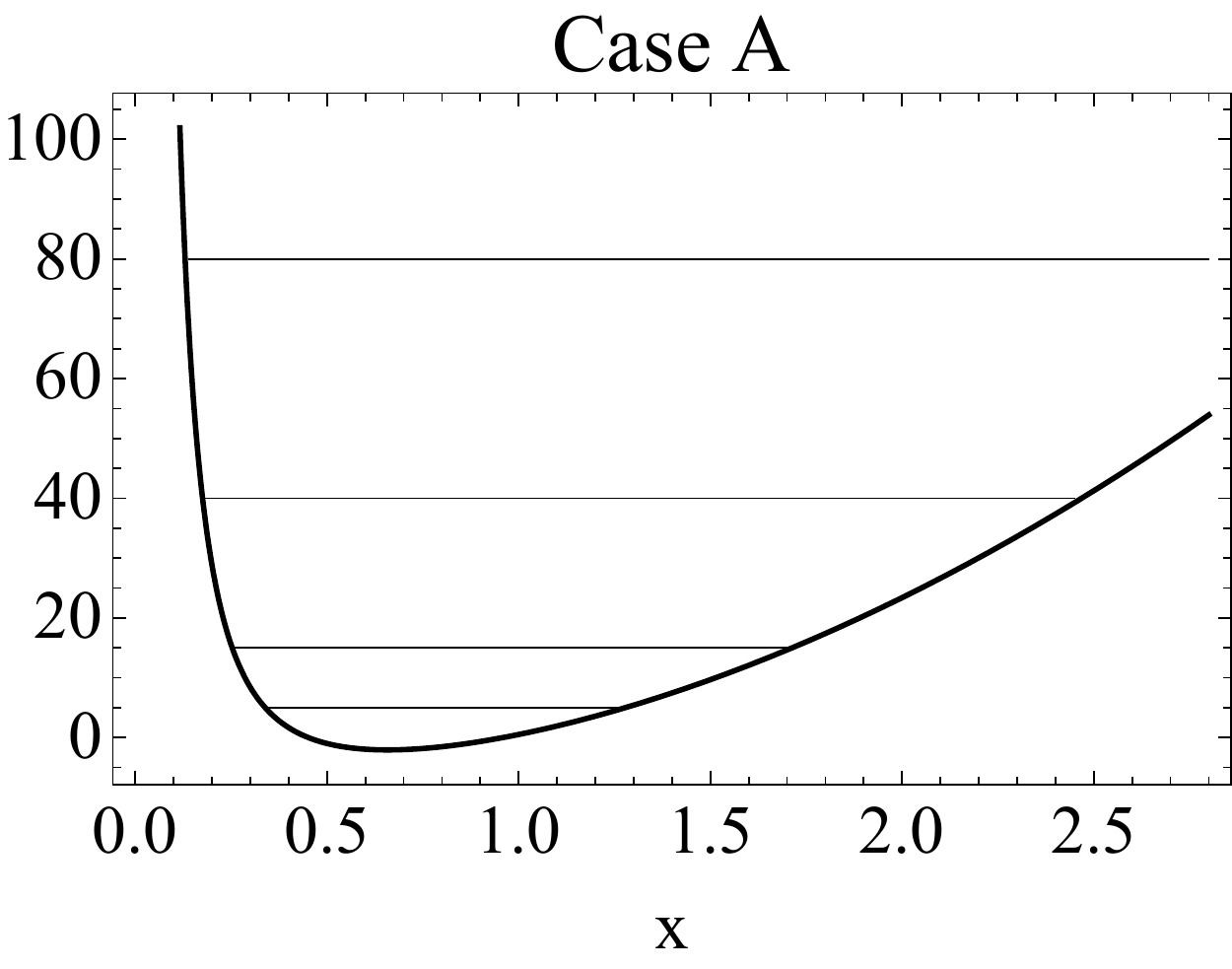}
\includegraphics[height=4cm]{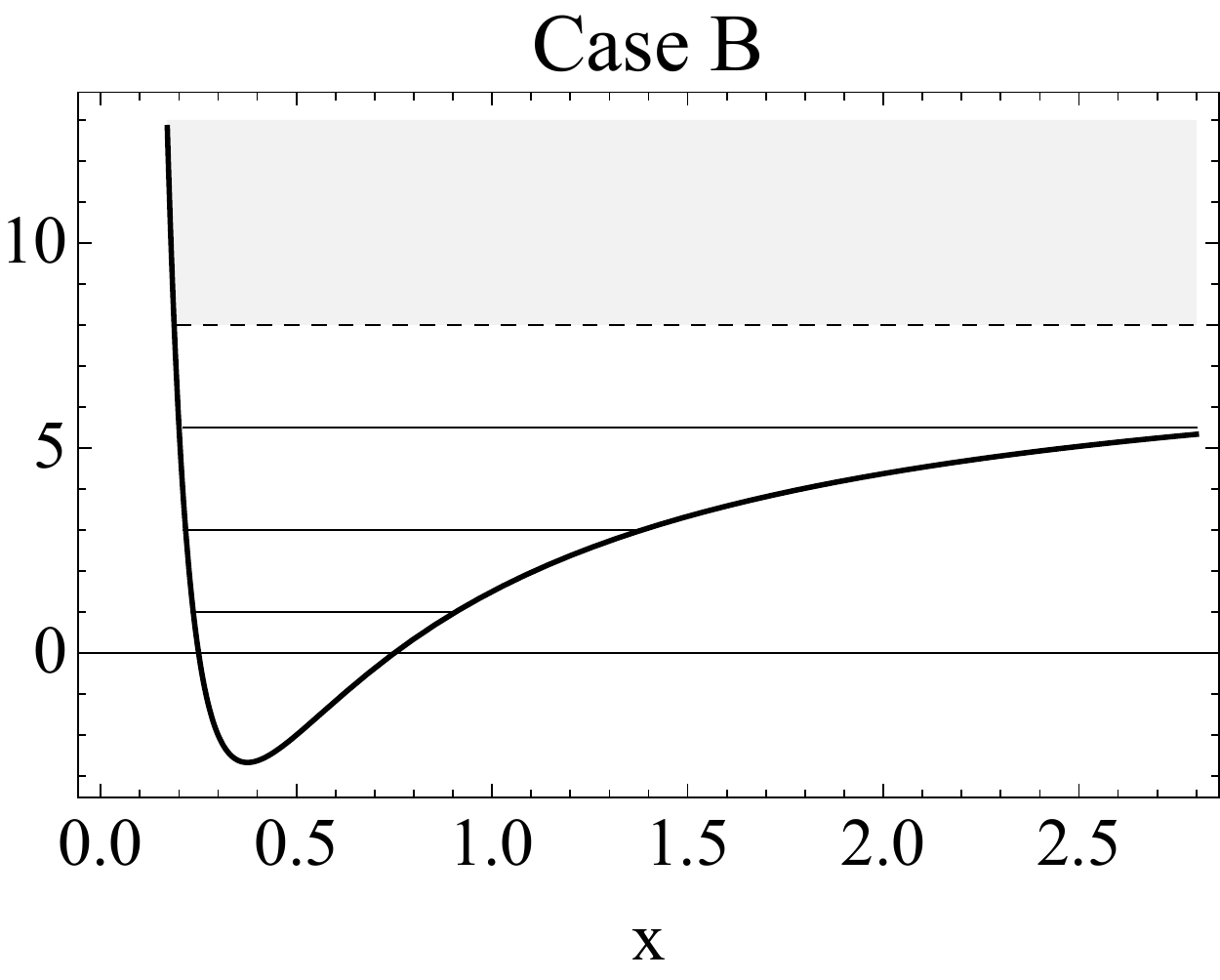}
\includegraphics[height=4cm]{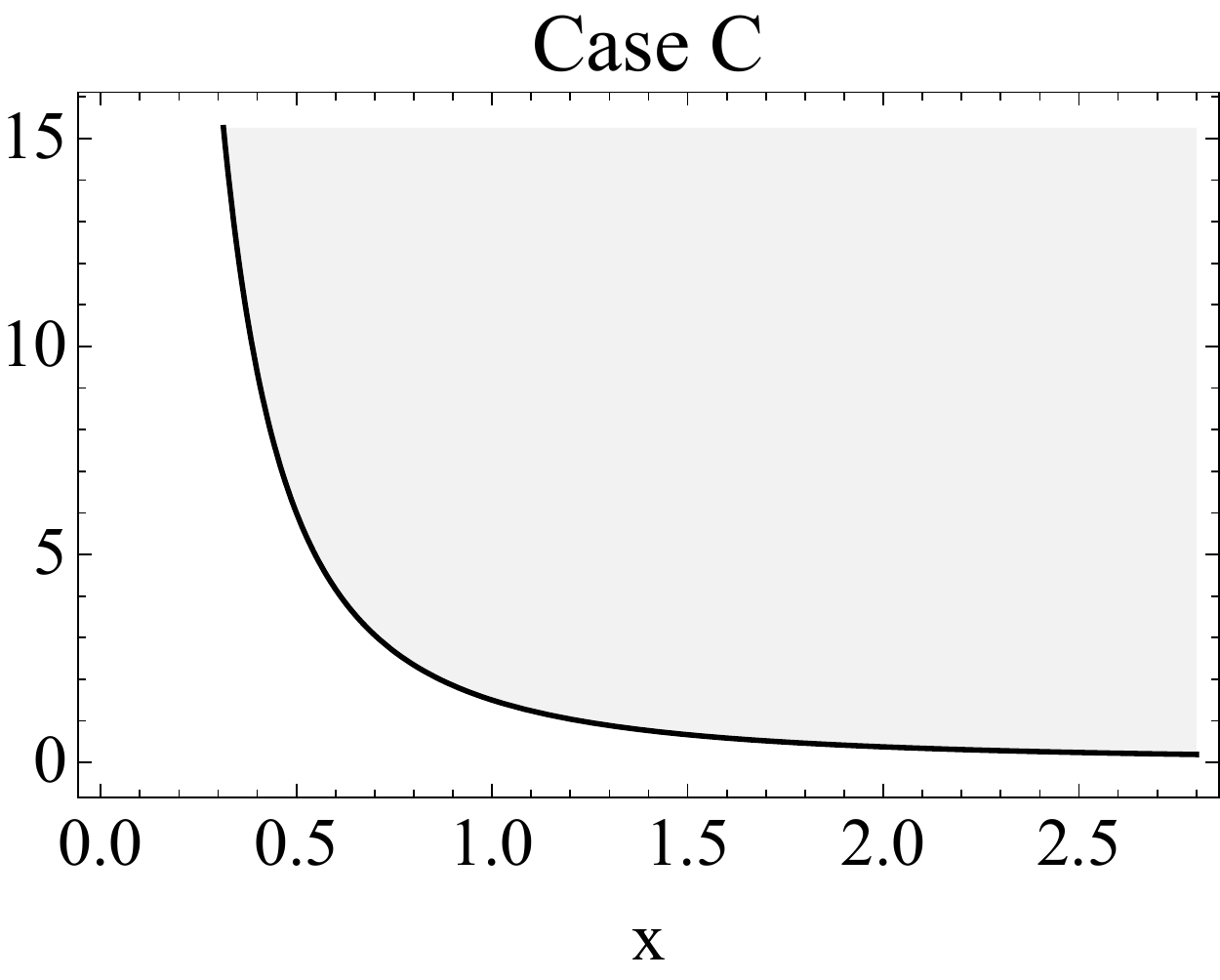}
\caption{Schr\"odinger potential in three qualitatively different cases on the half-line $\Omega=[0,\infty)$. The discrete spectrum is shown in case A (left panel),  the mixed spectrum is shown in case B (central panel), and the continuous spectrum is shown in case C (right panel).   Discrete eigenvalues are depicted by the lines and the continuum is represented by a gray shading.}
\label{fig:PlotSpectra}
\end{figure}
\end{center}

With the aid of Eq. \eqref{repa} we can reparametrize the Fokker-Planck equation  \eqref{FPEnew2} into the following Sturm-Liouville problem 
\be\label{Shreq}
- \lambda \, \psi_\lambda(x) = \mathcal{\tilde L}\, \psi_\lambda(x) = \left[\frsi\,  \p_x^2\, - \mathcal{V}_s(x)\right]\, \psi_\lambda(x) \, ,
\ee
which resembles an imaginary-time Schr\"odinger equation with potential
\ba\label{Schrpot}
\mathcal{V}_s(x) \equiv - \fr\, \left[d_x^2\, V(x) - \frac{1}{\s^2}\, (d_x\, V(x))^2 \right] = \fr \left[d_x\, f(x) + \frac{1}{\s^2}\, f(x)^2\right] \, . 
\ea
Again, the Sturm-Liouville theorem guarantees that the set of eigenfunctions, $\psi_\la(x)$, is orthonormal and complete, and all eigenvalues are real. In the case of a finite domain, the spectrum is an infinite set of discrete eigenvalues. However, if we take the limit of an infinite domain size the eigenvalue spectrum can either (a) remain a discrete set $\Lambda_d$, (b) become a continuum $\Lambda_c$, or (c) be a mix of the two $\Lambda= \Lambda_d + \Lambda_c$. Intuitively, as shown in Fig. \ref{fig:PlotSpectra}, the type of spectrum in the infinite domain case can be understood by depicting eigenvalues as lines of constant Schr\"odinger potential, Eq. \eqref{Schrpot}, or Schr\"odinger isopleths. Thus, a region of the spectrum wherein isopleths are bound at both extremes of the potential exhibits a discrete set of eigenvalues. 
Hence, since the operator $\mathcal{\tilde L}$ is self-adjoint, the spectrum in this region can be solved simply by imposing the boundary conditions on the generic solutions $\psi_\la(x) = \text{C}\, u_\la(x)$, where C is an integration constant.
The discrete set of solutions satisfying the boundary conditions will occur at values $\la_n$, $n=0,1,2,\cdots$, with $\lambda_{n=0}=0$ corresponding to the stationary solution, when it exists. Each discrete solution oscillates in the domain with $n$ zeros and is orthogonal to any solution with $m\neq n$, namely, 
\be
\int_\Omega\, dx\, \psi_n(x)\, \psi_m(x) \, = 0\, , \qquad m\neq n\, ,
\ee
where $n\equiv\lambda_n$ and $m\equiv\lambda_m$ are two eigenvalues. 
The integration constant is determined by normalization of the solution as
\be\label{normone}
\int_\Omega\, dx\, \text{C}\, u_n(x) = 1 \, ,
\ee
to satisfy the orthonormality relation
\be\label{normdisc}
\int_\Omega\, \,dx\,\psi_n(x)\, \psi_m(x) = \delta_{nm} \, ,
\ee
where $\delta_{nm}$ is the Kronecker delta.  In case A of Fig.\ref{fig:PlotSpectra} all isopleths are bound,  the entire spectrum is discrete, and the solution of the Fokker-Planck equation is given by Eq. \eqref{discretefull}.
When solved in a finite domain, a region of the spectrum wherein the potential bounds one or none of a line's extremes also exhibits a discrete set of eigenvalues. The spectrum is thus solved using the approach just described.   However, as shown in case C of Fig.\ref{fig:PlotSpectra}, when taking the boundaries to infinity, the set of lines becomes dense and tends towards a continuum.  Therefore, except special cases, there is a continuous set of solutions, $\psi^{con}$,  that oscillate asympotically at large $x$ with a finite constant amplitude. Oscillating functions cannot be normalized in the sense of Eq. \eqref{normdisc}, as the integral does not exist. That said, so long as they are real, they can be normalized in the sense of distributions to a Dirac $\delta$-function as
\be\label{dirac}
\int_\Omega \, dx\, \psi_{k_1}^{con}(x)\, \psi_{k_2}^{con}(x) = \delta(k_1-k_2) \, ,
\ee
where $k_1$ and $k_2$ are two eigenvalues in the continuous set. When the entire spectrum is a continuum, the sum in Eq. \eqref{discretefull} passes to an integral and the transition density is
\be\label{contfull}
\rho(x,t|x_0,t_0)^{con} = \sqrt{\frac{\rho_{st}(x)}{\rho_{st}(x_0)}}\, \int_{\Lambda}\, d\la\, \text{e}^{- \la\, (t-t_0)}\, \psi_\la^{con}(x_0)\, \psi_\la^{con}(x) \, .
\ee
Finally, at one (or both) of the boundaries approaching infinity, the potential may asymptotically approach a critical value $\la_{cr}$ that demarcates the continuous from the discrete region of the spectrum, as shown in panel B of Fig.\ref{fig:PlotSpectra}. Hence, we expect either a finite or an infinite  set of discrete eigenfunctions $\psi_{\la_n}^{dis}$, with eigenvalues $\la_n$, $n=0,1,2,\cdots$ for $\la<\la_{cr}$, and a continuous set, $\psi_{k}^{con}$, with eigenvalues $k\in[0,\infty)$, where $\la=\la_{cr} + k$. Thus, solution for the transition density in this case is
\ba\label{mix}
\rho(x,t|x_0,t_0)^{mix} = \sqrt{\frac{\rho_{st}(x)}{\rho_{st}(x_0)}} \left\{ 
\sum_{n=0}^{N}\, \text{e}^{- \la_n\, (t-t_0)}\, \psi_{\la_n}^{dis}(x_0)\, \psi_{\la_n}^{dis}(x)  \nonumber  \hspace{1em} \right. \\ \left.
+~\text{e}^{- \la_{cr}\, (t-t_0)}\, \int_{0}^{\infty}\, dk\, \text{e}^{- k\, (t-t_0)}\, \psi_k^{con}(x_0)\, \psi_k^{con}(x) 
\right\} \, ,
\ea
where $N$ is the highest discrete eigenvalue for a finite discrete set.
%
\subsection{\label{sec:NE}Normalization of oscillating functions}
As mentioned in \S \ref{sec:SE}, the oscillatory solutions cannot be normalized in the sense of Eq. \eqref{normone}. Nevertheless, the normalization provided by Eq. \eqref{dirac} does not provide an intuitive way to calculate the normalization constant. One approach is to solve the Sturm-Liouville problem in a finite domain of size $L$. The continuum of oscillating functions is then reduced to a discrete set of functions satisfying the boundary conditions of the problem. Then, the functions can be normalized in the sense of Eq. \eqref{normone}, after which we can take the limit $L\to\infty$ as in \cite{BKMH2013,BLIS1975}.  Here we adapt the approach introduced in the 1920s by Fues \cite{FUES1926, BSQM1957} to normalize quantum mechanical oscillatory eigenfunctions, which provides further insight into the structure of the Sturm-Liouville solutions. We extend the expansion of the solution shown in Eq. \eqref{eig1} to the case of a mixed spectrum as 
\be\label{eigmix}
\rho(x,t) = \sum_{n=0}^{N} \, \alpha_{n}(t_0)\, \text{e}^{- \la_n\, (t-t_0)}\, \phi_{n}^{dis}(x)  + \text{e}^{- \la_{cr}\, (t-t_0)}\, \int_0^{\infty}\,dk\, \text{e}^{- k\, (t-t_0)}\, a_k(t_0)\, \phi_k^{con}(x) \, ,
\ee
in which $N$ and $\la_{cr}$ are the same as in Eq.\eqref{mix},  $a_k(t_0)$ is the projection of the initial condition on the $k$-th eigenfunction, and $\phi(x) \propto \sqrt{\rho_{st}(x)}\, \psi(x)$. We then split the integral into a series of infinitesimal integrals viz., 
\be
\int_{0}^{\infty}\, dk\, \text{e}^{- k\, (t-t_0)}\, \psi_k^{con}(x) = \Lim{\Delta_n k \to 0} \, \sum_{n=1}^\infty\,  \text{e}^{- k_n\, (t-t_0)}\, \Psi_n(x) \, ,
\ee
where the $\Delta_n k$, $n=1,2,\cdots, \infty$, form an infinite discrete set of eigenvalue shells.  Here,  $k_n$ denotes the eigenvalue of the $n^{\text{th}}$ shell such that $\Lim{\Delta_n k \to 0} k_n = k$, 
and the functions $\Psi_n(x)$ are the \textit{eigendifferentials}, defined as the integrals of eigenfunctions over the $n^{\text{th}}$ shell;  
\be\label{eigendiff}
\Psi_n(x) = \int_{\Delta_n k}\,  dk' \,\psi_{k'}^{con}(x)\, .
\ee
Although a single eigenfunction $\psi_k^{con}(x)$ oscillates asymptotically in its argument, a linear combination of such functions does not, and hence the eigendifferentials decay at large $x$. Therefore, eigendifferentials play the same role in the continuum that the discrete eigenfunctions play in their spectral region. Consequently, discrete eigenfunctions are orthogonal to eigendifferentials, 
\be
\int_\Omega \, \psi^{dis}_n(x) \, \Psi_m(x)\, dx = 0 \, ,
\ee
for all $n$ and $m$. Thus, we can generalize the orthonormality definition to
\be\label{eigeortho}
\begin{cases}
\int_\Omega\, \psi^{dis}_n(x)\, \psi^{dis}_m(x)\, dx = \delta_{nm} \, , \\
\int_\Omega \, \psi^{dis}_n(x) \, \Psi_m(x)\, dx = 0 \, , \\
\frac{1}{\Delta_n k}\, \int_\Omega \, \Psi_n(x) \, \Psi_m(x)\, dx = \delta_{nm} \, .\\
\end{cases}
\ee
Determination of the factor $\Delta_n k$ in the orthonormalization rule for $\Psi_n(x)$ is given in Appendix \ref{AppendixFues}. The projection $a_k(t_0)$ in Eq. \eqref{eigmix} can then be defined as
\be
a_k(t_0) = \Lim{\Delta k\to0} \,  \frac{1}{\Delta k} \, \int_\Omega\, dx_0\,  \int_{k}^{k+\Delta k}\, dk\, \rho_{in}(x_0,t_0)\, \psi^{con}_k(x_0) \, .
\ee
Now, by using the eigendifferential definition of Eq. \eqref{eigendiff} in the third of Eqs. \eqref{eigeortho},  we obtain for $n=m$
\be
\frac{1}{\Delta_n k}\, \int_{\Delta_n k}\, dk_1\, \left\{ \int_\Omega\, dx \, \psi_{k_1}^{con}(x) \, \int_{\Delta_n k}\, dk_2\, \psi_{k_2}^{con}(x)\, \right\} = 1\, ,
\ee
where $\{k_1,k_2\} \in \Delta_nk$. The prefactor $(\Delta_n k)^{-1}$ insures that the above integral holds for any arbitrary shell size so long as
\be\label{contnormnew}
\int_\Omega \,  dx \, \psi_{k_1}^{con}(x) \, \left( \int_{\Delta_n k}\,dk_2\, \psi_{k_2}^{con}(x)\,  \right)  = \int_\Omega \, dx \, \text{C}(k_1)\, u(k_1, x) \, \left( \int_{\Delta_n k}\, dk_2\, \text{C}(k_2)\, u(k_2,x) \right) = 1\, ,
\ee
where $u(k,x)$ is the general solution of the Sturm-Liouville equation and $\text{C}(k)$ the normalization. Therefore, Eq. \eqref{contnormnew} generalizes Eq. \eqref{normone} to oscillatory solutions. 

Next we show that the normalization constant in Eq. \eqref{contnormnew} depends solely on the amplitude of the asymptotic oscillations of $\psi_k^{con}(x)$. Let us take the Schr\"odinger-like equation \eqref{Shreq} for two different eigenvalues $k_1$ and $k_2$, multiply them by $\psi^{con}_{k_2}$ and $\psi^{con}_{k_1}$ respectively, and then subtract them, which gives
\be\label{diveig}
\frsi\, \left\{\psi_{k_2}^{con}(x)\, \p_x^2\, \psi_{k_1}^{con}(x) - \psi_{k_1}^{con}(x)\, \p_x^2\, \psi_{k_2}^{con}(x) \right\}= (k_2 - k_1)\, \psi_{k_1}^{con}(x)\, \psi_{k_2}^{con}(x) \, .
\ee
Now, using Eq. \eqref{diveig} in Eq. \eqref{contnormnew}, and dividing by $(k_1-k_2)$, leads to
\begin{flalign}
&\int_\Omega \, dx\, \text{C}(k_1)\, u(k_1, x) \, \left( \int_{\Delta_n k}\,  dk_2\,  \text{C}(k_2)\, u(k_2,x)\right)\,   =  \qquad \nonumber \\
&  \int_{\Delta_n k}\, dk_2\, \left\{\int_\Omega\, dx \, \frac{\text{C}(k_1)\, \text{C}(k_2)}{k_2-k_1}\, \frsi\, [ u(k_2,x)\, \p_x^2\, u(k_1,x) 
 - u(k_1,x)\, \p_x^2\, u(k_2,x) ]\, \right\} = 1\, .
\end{flalign}
We integrate by parts in a finite domain with the boundary terms defined at infinity, and consider two domains; (i) $\Lim{x_{out}\to\infty}[-x_{out},x_{out}] = (-\infty, +\infty)$, and (ii) the half-line $\Lim{x_{out}\to\infty}[x_{in},x_{out}] = [x_{in}, \infty)$, with the density vanishing at both boundaries. In case (i) we find
\be\label{asas}
\Lim{x_{out}\to\infty}\, \int_{\Delta_n k}\, dk_2\, \left\{\frac{\text{C}(k_1)\, \text{C}(k_2)}{k_2-k_1}\, \frsi\, \Big[ u(k_2,x)\, \p_x\, u(k_1,x) - u(k_1,x)\, \p_x \, u(k_2,x) \Big]_{-x_{out}}^{x_{out}}\,\right\} = 1\, ,
\ee
where the squared parenthesis denote the difference in the value of the function at the extremes viz., $[f(x)]_{{-x_{out}}}^{{x_{out}}}= f({x_{out}}) - f({-x_{out}})$.
With the exception of a few cases, such as the Bessel process with constant drift (discussed in \S \ref{sec:BE}), the asymptotic behavior of the oscillating eigenfunctions satisfies a Helmholtz equation. By taking the limit $x\to\infty$ in the Schr\"odinger equation \eqref{Shreq}, and hence considering $\Lim{x\to\infty}\, \mathcal V_s(x) = \la_{cr}$, the asymptotic behavior of the solution $\tilde u(k,x)$ satisfies
\be
\frsi\, \partial_x^2\, \tilde u(k,x) + (\la - \la_{cr})\, \tilde u(k,x) = \frsi\, \partial_x^2\, \tilde u(k,x) + k\, \tilde u(k,x)= 0\, ,
\ee
so that the generic asymptotic behavior can be written as 
\be\label{solas}
\tilde u_1(k,x) = \Aa(k)\, \cos\left(\frac{\sqrt{2~k}\, x}{\s}\right)\, \qquad \text{or} \qquad  \tilde u_2(k,x) = \Bb(k)\, \sin\left(\frac{\sqrt{2~k}\, x}{\s}\right) \, ,
\ee
with amplitudes $\Aa(k)$ and $\Bb(k)$.  At this stage we define one eigendifferential for each class of free solutions, whereas Fues defines eigendifferentials of linear combinations of solutions, which do not reproduce the solution of the Brownian motion problem discussed in \S \ref{sec:RBM}, and additionally the associated integrals are very complicated. 

Now, substituting the cosine in Eq. \eqref{solas} into Eq. \eqref{asas} the quantity in the square brackets in the integrand becomes 
\ba
\label{eqe1}
&&\Big[ \tilde u_1(k_2,x)\, \p_x\, \tilde u_1(k_1,x) - \tilde u_1(k_1,x)\, \p_x \, \tilde u_1(k_2,x) \Big]_{-x_{out}}^{x_{out}} \nonumber \\
&&=\Big[w_1\,\left(- \mathcal A_1\, \sin(w_1\, x) \right)\, \left(\Aa_2\,\cos(w_2\, x)\, \right) - w_2\,\left(- \mathcal A_2\, \sin(w_2\, x) \right)\, \left(\Aa_1\,\cos(w_1\, x)\, \right) \Big]_{-x_{out}}^{x_{out}},
\ea
where $w_i \equiv \sqrt{2\, k_i}/\s$, $i\in\{1,2\}$, and $\Aa_i \equiv \Aa(k_i)$\footnote{Note that using the sine in Eq. \eqref{solas} gives the same expression as the cosine, but leads to an opposite sign in the second row of Eq. \eqref{eqeq}, which eventually vanishes.}.
Using $\cos(a) \sin(b) = \fr [\sin(a+b) - \sin(a-b)]$, the previous equation becomes
\ba\label{eqeq}
&&\Big[ \tilde u_1(k_2,x)\, \p_x\, \tilde u_1(k_1,x) - \tilde u_1(k_1,x)\, \p_x \, \tilde u_1(k_2,x) \Big]_{-x_{out}}^{x_{out}} \nonumber \\
&&=\Big[ 
\sin((w_2 - w_1)\, x)\, (\Aa_2\, \Aa_1)\, \frac{(w_2 + w_1)}{2} 
+ \sin((w_2 + w_1)\, x)\, (\Aa_2\, \Aa_1)\, \frac{(w_2 - w_1)}{2} \Big]_{-x_{out}}^{x_{out}} \,. 
\ea
In order to integrate Eq. \eqref{asas} over $k_2$ we note that the amplitudes $\Aa_i$ are very slowly varying functions of $k$, and thus we treat them as constants with respect to integration without introduction of significant error.  The precision of this procedure must be thoroughly assessed numerically throughout the parameter space, making the method difficult to use. 

\begin{center}
\begin{figure}[ht]
\includegraphics[height=3.8cm]{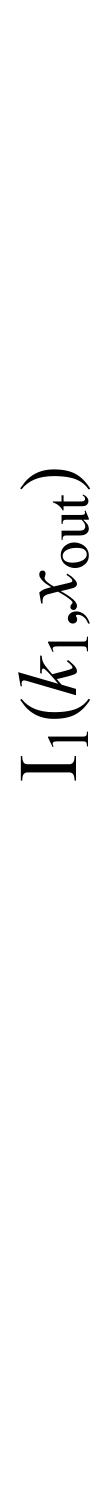}
\includegraphics[height=3.8cm]{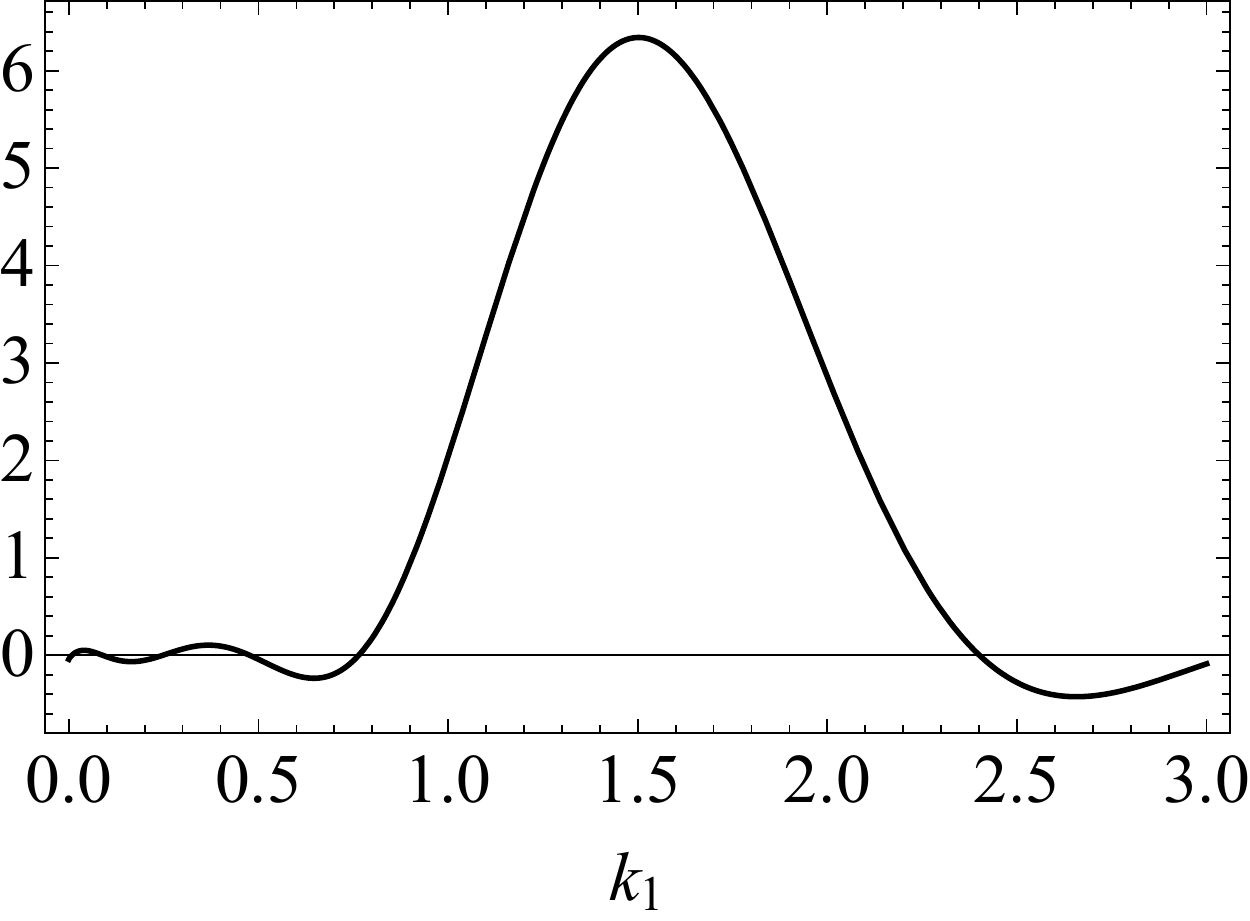}
\includegraphics[height=3.8cm]{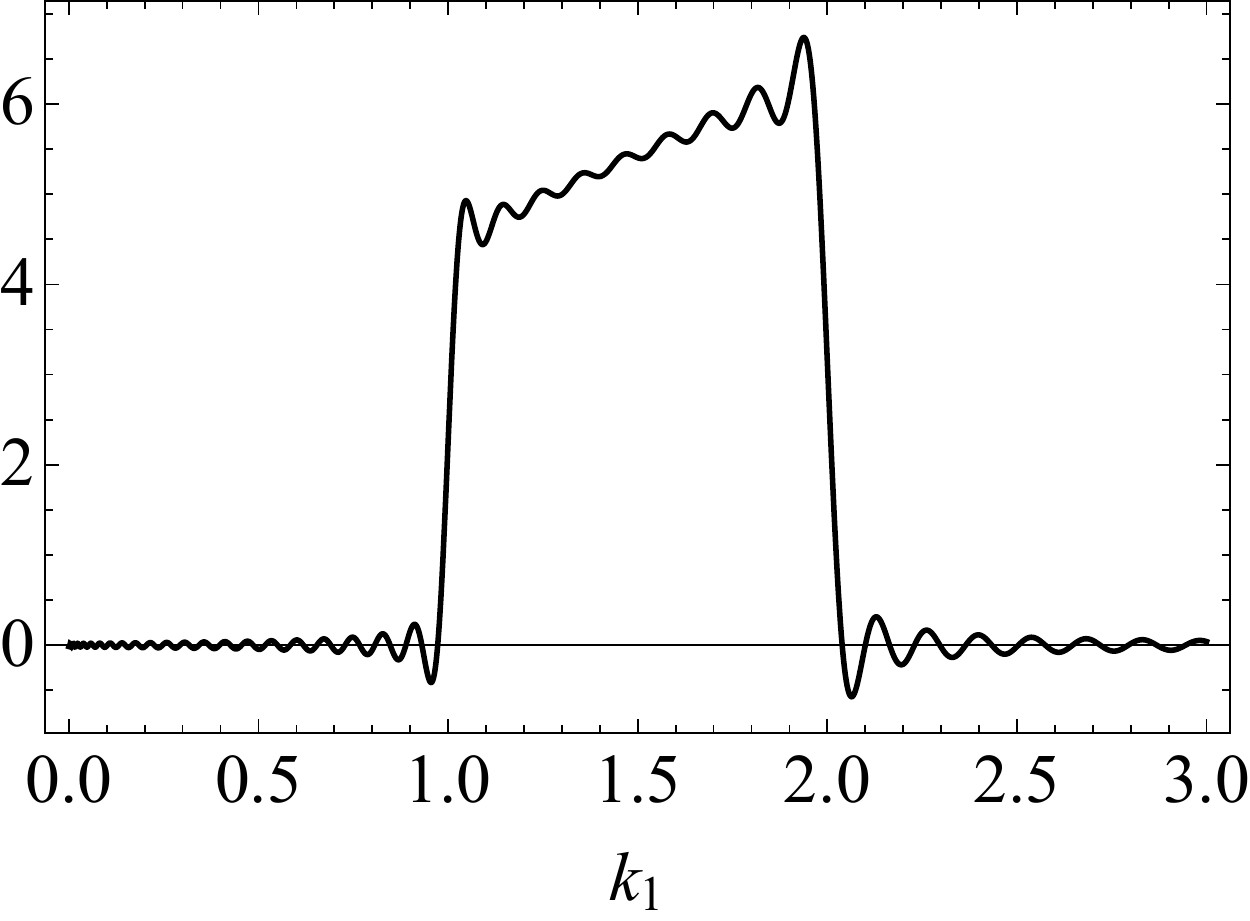}
\includegraphics[height=3.8cm]{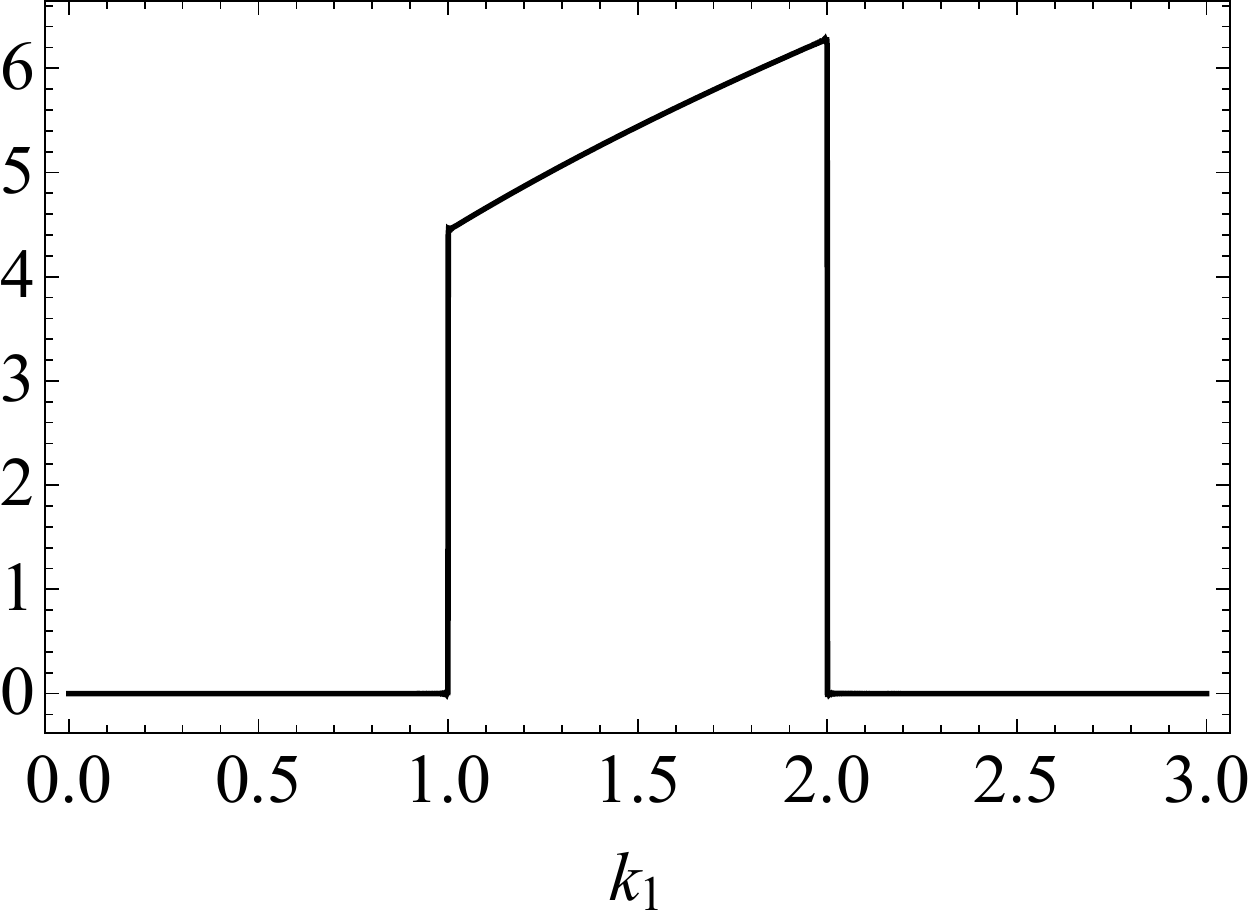}
\caption{The function $I_1(k_1,x_{out})$ for $x_{out} = 10$ (left panel), $x_{out} = 10^2$ (central panel), and $x_{out} = 10^6$ (right panel). The other parameters are $k_{in}=1, k_{out}=2,$ and $\s=1$.}
\label{fig:PlotTheta}
\end{figure}
\end{center}

Because Eq. \eqref{eqeq} contains only odd functions so that $\sin(x) = - \sin(-x)$ and $\sin(0) = 0$, we can apply the rule $[\, ]_{-x_{out}}^{x_{out}} \to 2\, [\, ]_0^{x_{out}}$, obtaining
\be
\Lim{x_{out}\to\infty}\, 2\, \text{C}(k_1)\, \text{C}(k_2)\, \frsi\,  (\Aa_2\, \Aa_1)\,[ I_1(k_1, x_{out}) + I_2(k_1, x_{out})]\, = 1 \, ,
\ee
where
\begin{align}
\label{int1sin}
I_1(k_1,x_{out})  = & \int_{k_{in}}^{k_{out}}\, dk_2\, \frac{(w_2 + w_1)}{2}\, \frac{\sin((w_2 - w_1)\, x_{out})}{k_2 - k_1} \nonumber \\ 
 = &\frac{\cos\left(\frii\, \left(\sqrt{k_1} - \sqrt{k_{in}}\right)\, x_{out}\right)}{x_{out}} - \frac{\cos\left(\frii\, \left(\sqrt{k_1} - \sqrt{k_{out}}\right)\, x_{out}\right)}{x_{out}} \nonumber\\
 + & \frii\, \sqrt{k_1}\, \text{Si}\left( \frii \left( \sqrt{k_1} - \sqrt{k_{in}} \right)\, x_{out}\right) - \frii\, \sqrt{k_1}\, \text{Si}\left( \frii \left( \sqrt{k_1} - \sqrt{k_{out}} \right)\, x_{out}\right) \, , \nonumber\\
 \text{and} \nonumber\\
I_2(k_1,x_{out}) = &\int_{k_{in}}^{k_{out}}\, dk_2\, \frac{(w_2 - w_1)}{2}\, \frac{\sin((w_2 + w_1)\, x_{out})}{k_2-k_1} \nonumber \\
= &\frac{\cos\left(\frii\, \left(\sqrt{k_1} + \sqrt{k_{in}}\right)\, x_{out}\right)}{x_{out}} - \frac{\cos\left(\frii\, \left(\sqrt{k_1} + \sqrt{k_{out}}\right)\, x_{out}\right)}{x_{out}} \nonumber\\
+ &\frii\, \sqrt{k_1}\, \text{Si}\left( \frii \left( \sqrt{k_1} + \sqrt{k_{in}} \right)\, x_{out}\right) - \frii\, \sqrt{k_1}\, \text{Si}\left( \frii \left( \sqrt{k_1} + \sqrt{k_{out}} \right)\, x_{out}\right) \, , \nonumber\\
\end{align}
where $\Delta_n k = [k_{in}, k_{out}]$, and $\text{Si}(x)$ is the sine integral function. By taking the boundary to infinity, the cosines in Eq. \eqref{int1sin} are suppressed while the sine integrals tend to a Heaviside theta function, 
\be
\Lim{x_{out}\to\infty} \text{Si}(x_{out}\, (a+b)) = \pi\, \left(\Theta(a+b) - \fr\right) \, ,
\ee
where $\Theta(a+b)$ is zero (one) for $b<-a$ ($b>-a$).
Due to the two Heaviside functions,  for $k_1 \in \Delta_n k$ and $k_1 \in - \Delta_n k$, the integrals $I_1(k_1,x_{out})$ and $I_2(k_1,x_{out})$ are equal to $\pi\, \sqrt{k_1\, 2}/\s$ and zero otherwise. Therefore, upon taking the infinite domain limit, the integral $I_2$ vanishes and the integral $I_1$ takes on values solely inside the shell, as depicted in Fig. \ref{fig:PlotTheta}. Therefore, the orthonormalization rule Eq. \eqref{eigeortho} is verified. Finally, we reparametrize the integral around $k_1=k_2$ as
\ba
\Lim{x_{out}\to\infty}\,2\, \text{C}(k_1)^2\, \frsi\,
\Aa_1^2\, \int_{k_1-\eta}^{k_1+\eta}\, dk_2\, \frac{(w_2 + w_1)}{2}\, \frac{\sin[(w_2 - w_1)\, x_{out}]}{k_2-k_1} & = \nonumber \\  2 \, \text{C}(k_1)^2\, \frsi\,
\Aa_1^2\, \frii\, \sqrt{k_1}\, \pi\,  & = 1 \, ,
\ea
from which the squared normalization is
\be\label{normful1}
\text{C}(k)_\text{full}^2 = \frac{1}{\sqrt{2\, k}\, \s\, \pi\, \, \Aa(k)^2}\,  .
\ee
The calculation for the half domain case proceeds analogously and, due to the different extremes in Eq. \eqref{eqeq}, differs only by a factor two; 
\be\label{normhalf1}
\text{C}(k)_\text{half}^2 = \frac{\sqrt{2}}{\sqrt{k}\,\s \, \pi\, \Aa(k)^2}\, .
\ee
%
\vspace{-0.2in}
\section{\label{sec:RBM}Brownian motion with constant drift}
In this section we demonstrate the technique by solving the Fokker-Planck equation for Brownian motion with a constant drift function $f(x) = a$
\be\label{RBMeq}
\p_t\, \rho(x,t) = - a\, \p_x \, \rho(x,t) + \frsi \, \p_x^2\, \rho(x,t) \, ,
\ee
and natural boundary conditions, $\rho(x=\pm\infty,t)=0$.  
Regardless of whether we have positive  ($a>0$) or  negative ($a<0$) deterministic drift, the potential $V(x) = - a\, x$ will be unbounded at $x=\pm\infty$. Therefore, the stationary solution 
\be
\rho_{st}(x) = \frac{\mathcal N}{\s^2}\text{e}^{-\frs\, V(x)} = \frac{\mathcal N}{\s^2}\text{e}^{\frac{2\, a}{\s^2}\, x} \, ,\\
\ee
is not normalizable, and the process is intrinsically transient.
We can solve the Fokker-Planck equation by solving the associated Schr\"odinger eigenvalue problem, Eq. \eqref{Shreq}, with constant Schr\"odinger potential
\be
\mathcal V_s(x) = \frac{a^2}{2\, \s^2}\, 
\ee
for which the Sturm-Lioville problem has only a continuum of eigenvalues. Since the stationary solution is not normalizable the eigenvalue $\la=0$ does not belong to the spectrum, which consists only of oscillating solutions. 

\begin{center}
\begin{figure}[ht]
\includegraphics[height=3.8cm]{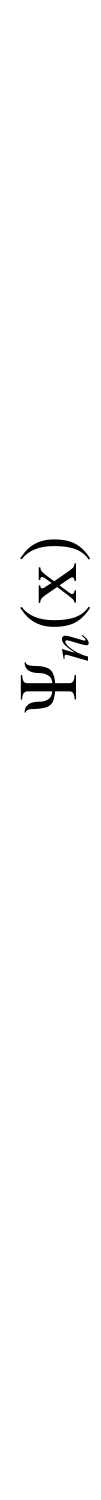}
\includegraphics[height=3.8cm]{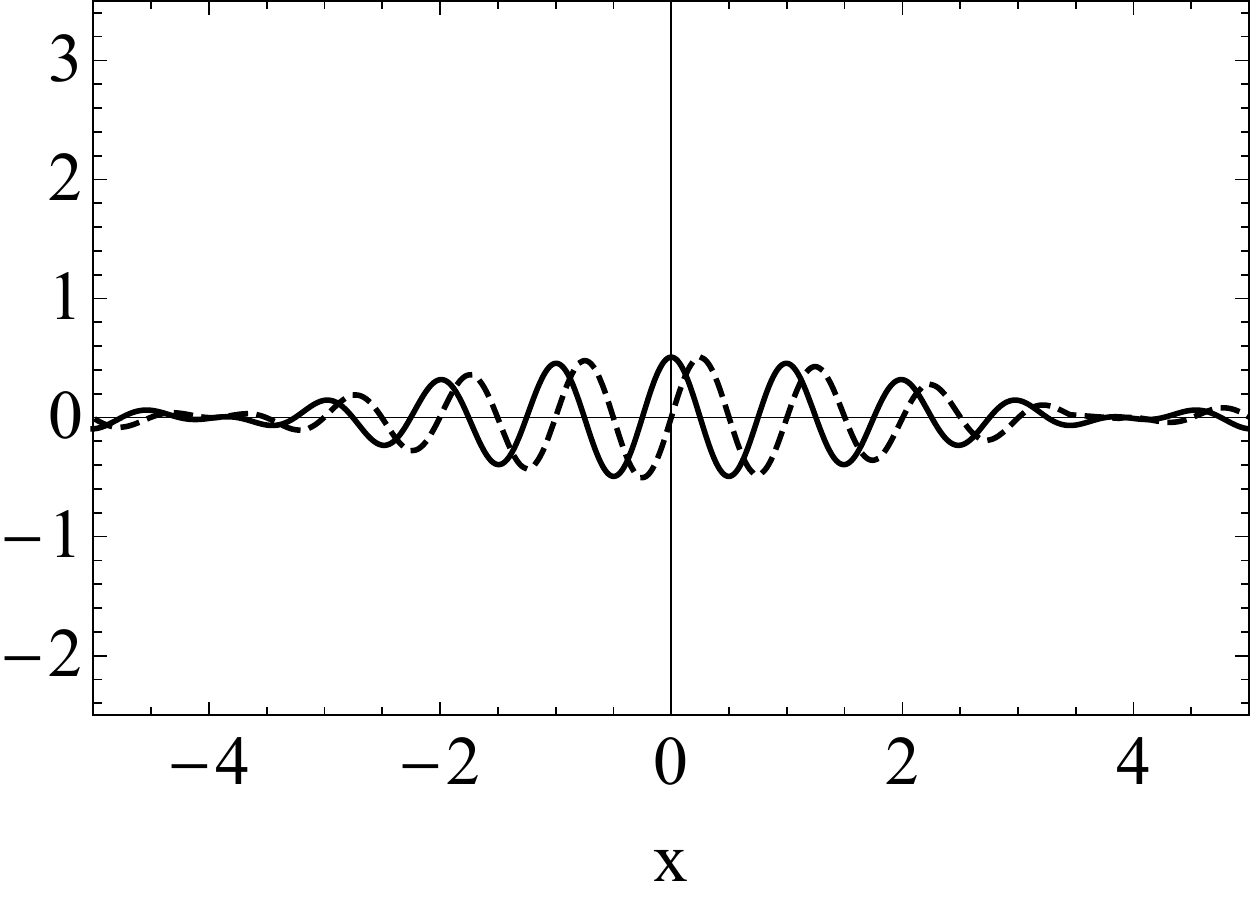}
\includegraphics[height=3.8cm]{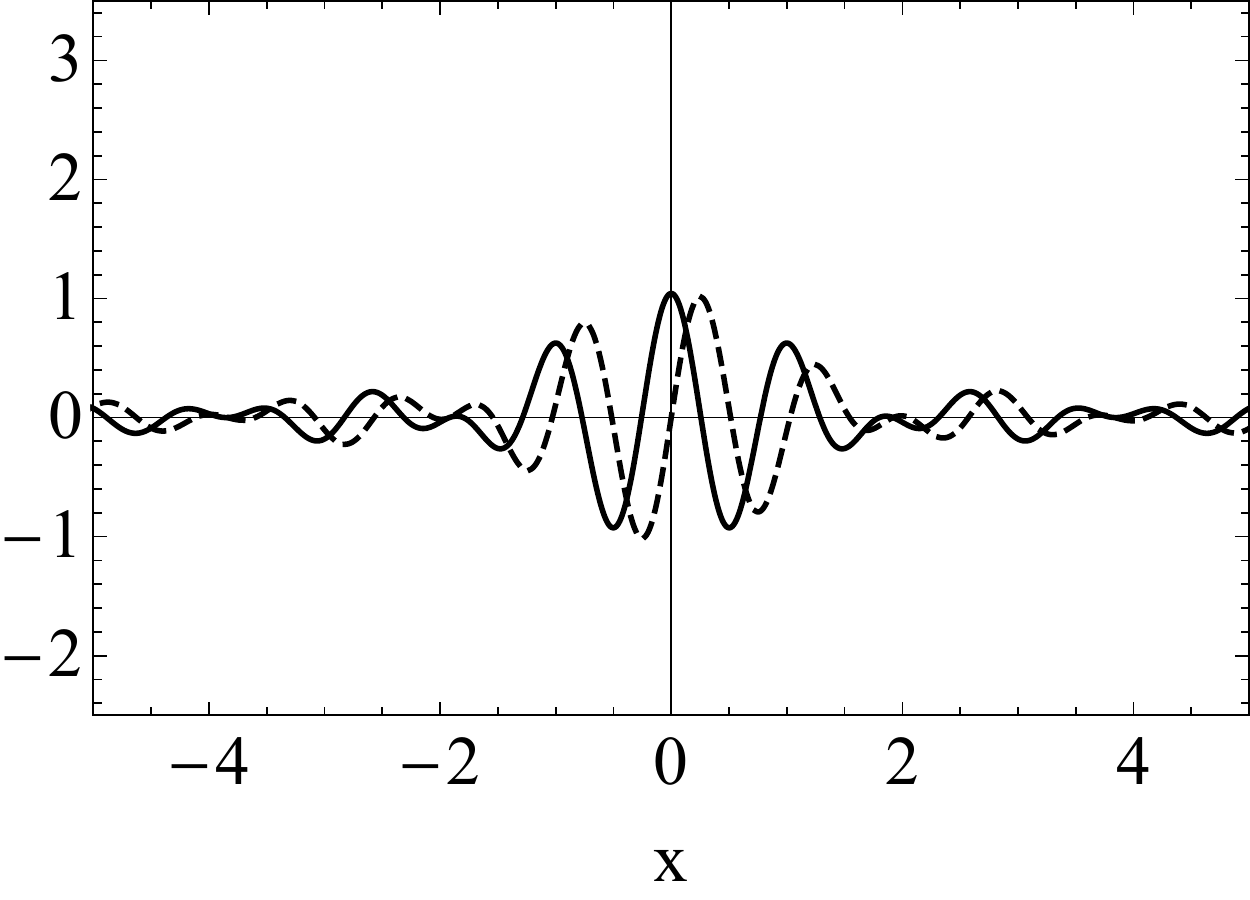}
\includegraphics[height=3.8cm]{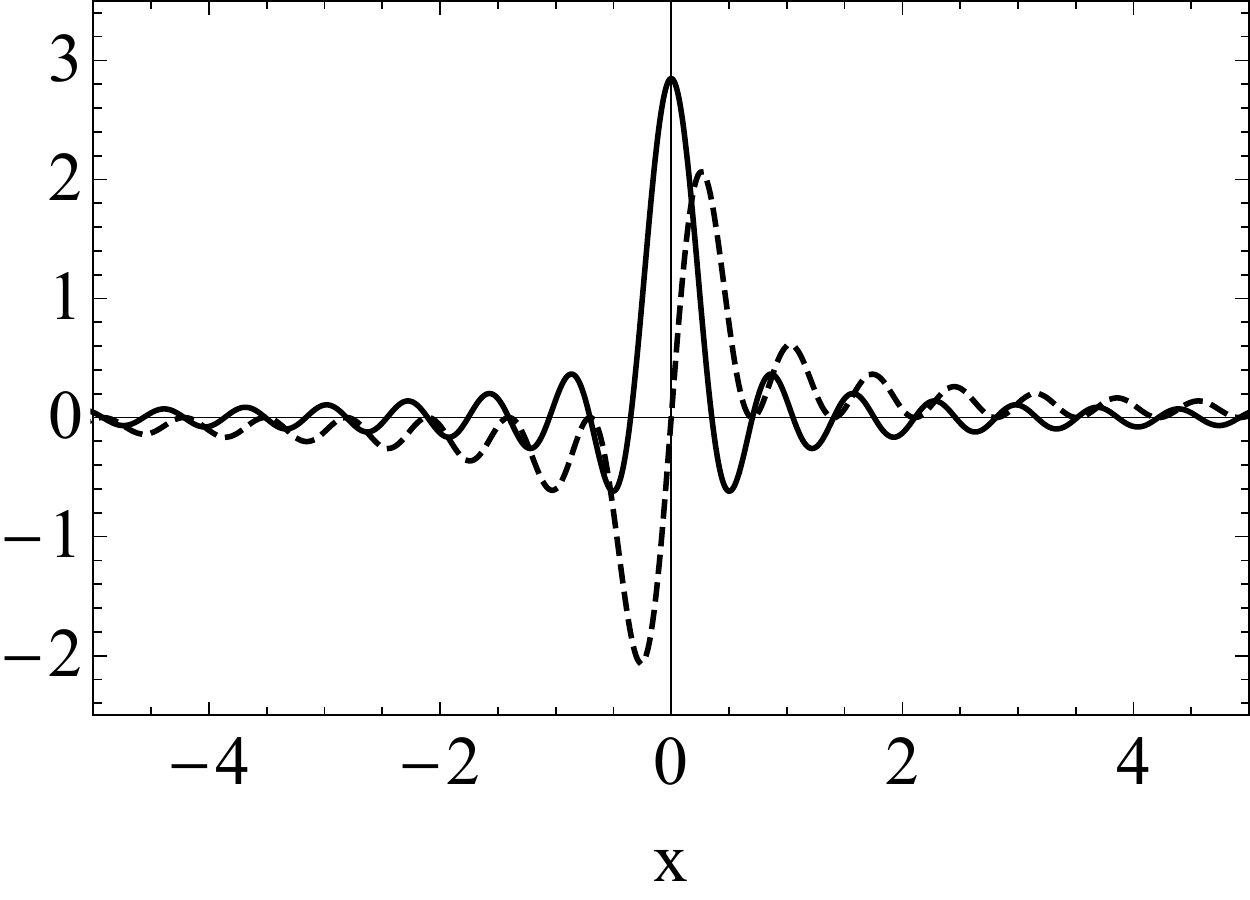}
\caption{The eigendifferentials $\Psi_n^{cos}(x)$ (solid line) and $\Psi_n^{sin}(x)$ (dashed line) for $k_n = 20, \s = 1$, $\eta = 5$ (left panel), $\eta = 10$ (central panel) and $\eta = 20$ (right panel).}
\label{fig:PlotEigend}
\end{figure}
\end{center}

We begin by writing the general solution $u_\la(x)$ of the Schr\"odinger equation,  
\be
u_\la(x) = \text{C}_1\, \exp\left\{\frac{x\, \sqrt{a^2 - 2\, \la\, \s^2}}{\s^2}\right\} + \text{C}_2\, \exp\left\{-\frac{x\, \sqrt{a^2 - 2\, \la\, \s^2}}{\s^2}\right\} \, ,
\ee 
with $\text{C}_1$ and $\text{C}_2$ the integration constants. We exclude the case where $\la < a^2/(2\, \s^2)$ because both solutions diverge at one boundary and vanish at the other. For $\la \geq a^2/(2\, \s^2)$ both the linearly independent solutions are imaginary exponentials, so that sines and cosines are solutions viz., 
\be\label{gensol}
u_{\la}(x) = \text{C}_1\, \cos\left(\frac{\sqrt{2~k}~x}{\s}\right) + \text{C}_2\, \sin\left(\frac{\sqrt{2~k}~x}{\s}\right) \, ,
\ee
with $\la = a^2/(2\, \s^2) + k$. Note that these oscillatory eigenfunctions of Brownian motion have the same asymptotic behavior as Eq. \eqref{solas}, with amplitudes $\Aa=\Bb =1$. Therefore, the norm is given by Eq. \eqref{normful1} and the normalized eigenfunctions are
\be\label{deffull}
\psi^{sin}_k(x) = \mathcal N\, \sin\left(\frac{\sqrt{2~k}~x}{\s}\right) \qquad \text{and} \qquad \psi^{cos}_k(x) = \mathcal N\, \cos\left(\frac{\sqrt{2~k}~x}{\s}\right)\,;  \qquad \mathcal N = \frac{1}{(\sqrt{2\, k}\,\pi\, \s)^{\fr}} \, . 
\ee
The solution can then be written as a sum of two classes of eigendifferentials $\Psi_n^{cos}(x)$ and $\Psi_n^{sin}(x)$ as
\be
\rho(x,t|x_0,t_0) = \Lim{\eta\to0}\, \int_{-\infty}^{\infty}\, dx_0\, \sum_{n=1}^{\infty}\, \text{e}^{- k_n\, (t-t_0)}\, \left\{\Psi_n^{cos}(x)\, \Psi_n^{cos}(x_0)\, + \Psi_n^{sin}(x)\, \Psi_n^{sin}(x_0) \right\} \,  ,
\ee
where
\begin{flalign}
\Psi_n^{cos}(k) & = \int_{k_n-\eta}^{k_n+\eta}\, dk' \, \frac{1}{\sqrt{2\, k'}\, \pi\, \s}\, \cos \left(\frac{\sqrt{2~k'}}{\s}\, x\right) \nonumber\\ & = \frac{1}{\pi\, x}\, \left[- \sin\left( \frac{\sqrt{2}}{\s}\, \sqrt{k-\eta}\, x\right) + \sin\left( \frac{\sqrt{2}}{\s}\, \sqrt{k+\eta}\, x\right) \, \right] \, ,
\end{flalign}
\text{and} 
\begin{flalign}
\Psi_n^{sin}(k) &= \int_{k_n-\eta}^{k_n+\eta}\, dk' \, \frac{1}{\sqrt{2\, k'}\, \pi\, \s}\, \sin \left(\frac{\sqrt{2~k'}}{\s}\, x\right) \nonumber\\ &= \frac{1}{\pi\, x}\, \left[\,\cos\left( \frac{\sqrt{2}}{\s}\, \sqrt{k-\eta}\, x\right) - \cos\left( \frac{\sqrt{2}}{\s}\, \sqrt{k+\eta}\, x\right) \, \right] \, .
\end{flalign}
Here the shell size range is $\Delta_n k = [k_n-\eta,k_n+\eta]$ and the two sets of eigendifferentials are shown in Fig. \ref{fig:PlotEigend} for different shell sizes.
Finally, the solution of the process can be determined by substituting Eq. \eqref{deffull} into Eq. \eqref{contfull}, which gives 
\begin{flalign}
\rho(x,t|x_0,t_0) &= \int_{-\infty}^{\infty}\, dx_0\, \sqrt{\frac{\rho_{st}(x)}{\rho_{st}(x_0)}}\, \text{e}^{- \frac{a^2}{2\, \s^2} \, (t-t_0)}\, \int_0^{\infty}\, dk\,  \text{e}^{-k\, (t-t_0)} \left\{ \psi^{sin}_k(x)\, \psi^{sin}_k(x_0) + \psi^{cos}_k(x) \,  \psi^{cos}_k(x_0) \right\} \, \nonumber\\
&= \int_{-\infty}^{\infty}\, dx_0\, \sqrt{\text{e}^{\frac{2\, a\, (x-x_0)}{\s^2}}} \, \int_0^{\infty}\, dk \, \frac{1}{\sqrt{2\, k}\, \pi \, \s} \, \text{e}^{- \left[\left(\frac{a^2}{2\, \s^2} + k\right) \, (t-t_0)\right]}\, \nonumber\\
&\times \left\{\cos \left(\frac{\sqrt{2~k}~x}{\s}\right)\, \cos \left(\frac{\sqrt{2~k}~x_0}{\s}\right)  + \sin \left(\frac{\sqrt{2~k}~x}{\s}\right)\, \sin \left(\frac{\sqrt{2~k}~x_0}{\s}\right)\right\} \, .
\end{flalign}
By using the trigonometric relation 
\be
\cos(\alpha\, x)\, \cos(\alpha\, x_0)\, + \sin(\alpha\, x)\, \sin(\alpha\, x_0) = \cos[\alpha\, (x-x_0)] \, ,
\ee
and
\be
\int_0^\infty\, dk\, \frac{1}{\sqrt{k}}\, \text{e}^{-\beta\, k}\, \cos(\alpha\, \sqrt{k}) = \sqrt{\frac{\pi}{\beta}}\, \text{e}^{- \frac{\alpha^2}{4\, \beta}}\, ,
\ee
we obtain the well-known Gaussian distribution; 
\be
\rho(x,t|x_0,t_0) = \frac{1}{\sqrt{2\, \pi\, t\, \s^2}} \, \text{e}^{-\frac{(x - x_0 - a\, t)^2}{2\, \s^2\, t}}\, ,
\ee
with a broadening enhanced by the constant deterministic drift force characterized by $a$.  

To determine the solution in the positive half domain, the general solution, Eq. \eqref{gensol}, must satisfy the boundary condition at the origin. Hence, we discard the cosines, and the normalization is now provided by Eq. \eqref{normhalf1}. However, because the remaining set of sines have a non-vanishing derivative at the origin, we have an absorbing boundary condition there and a natural boundary condition at infinity, giving the solution as 
\begin{flalign}
\rho(x,t|x_0,t_0)_\text{half} &= \int_{0}^{\infty}\, dx_0\, \sqrt{\text{e}^{\frac{2\, a\, (x-x_0)}{\s^2}}}\,  \int_0^{\infty}\, dk \, \frac{\sqrt{2}}{\sqrt{k}\, \pi \,\s}\, \text{e}^{- \left(\frac{a^2}{2\, \s^2} + k\right) \, (t-t_0)}\,
\nonumber\\ 
& \times \left\{\sin \left(\frac{\sqrt{2~k}\, x}{\s}\right)\, \sin \left(\frac{\sqrt{2~k}\, x_0}{\s}\right)\right\} \, , 
\end{flalign}
which can be simplified using the integral relation
\be
\int_0^\infty\, dk\, \frac{1}{\sqrt{k}}\, \text{e}^{-\beta\, k}\, \sin(\alpha_1\, \sqrt{k}) \, \sin(\alpha_2\, \sqrt{k}) = \frac{\sqrt{\pi}\,  \left(\text{e}^{-\frac{(\alpha_1 - \alpha_2)^2}{4\, \beta }} - \text{e}^{-\frac{(\alpha_1 + \alpha_2)^2}{4\, \beta }}\right)}{2\, \sqrt{\beta}}\, ,
\ee
finally leading to the solution of Eq. \eqref{RBMeq} as, 
\be\label{solmethod}
\rho(x,t|x_0,t_0)_\text{half} = \frac{1}{\sqrt{2\, \pi\, t\, \s^2 }} \, \left[ \text{e}^{-\frac{(x- x_0 - a\, t)^2}{2\, \s^2\,  t}} - \text{e}^{\left(- \frac{(x + x_0 - a\, t)^2}{2\, \s^2\, t} - \frac{2\, a\, x_0}{\s^2}\right)} \right] \, .
\ee
We conclude this section by emphasizing that reproducing known results acts as the ideal test bed for the technique.  Note that Eq. \eqref{solmethod} can be obtained by employing the method of images and we also direct the reader to the work of Linetsky \cite{VALI2005} regarding the spectral representation of Brownian motion with a reflecting boundary condition at the origin (see also Abate and Whitt \cite{ABWH1987a, ABWH1987b}) .
%
\vspace{-0.2in}
\section{\label{sec:BE}Bessel process with constant drift}
Now we treat the Bessel process with constant negative drift, described a Fokker-Planck equation with potential $V(x) = - b\, \ln(x) - a\, x$, $a<0$ and $b>0$, with a natural boundary condition at infinity and a reflecting (i.e., no-flux) boundary at the origin that implies $\phi(x=0)=0$. The process has the normalizable stationary solution, 
\ba\label{rhos}
\rho_{st}(x) = \frac{\mathcal N}{\s^2}\, \text{e}^{- \frs\, V(x)} =  \frac{\mathcal N}{\s^2}\, x^{\frac{2\, b}{\s^2}}\, \text{e}^{\frac{2\, a}{\s^2}\, x}\, ,
\ea
where 
\ba\label{statnorm}
\mathcal N =  \int_0^\infty\,dx\, \rho_{st}(x) = \left[ b\, 4^{-\frac{b}{\s^2}}\, \s^{\frac{4\, b}{\s^2}-2}\, (-a)^{-\frac{2\, b}{\s^2}-1}\, \Gamma\left(\frac{2\,b}{\s^2}\right) \right]^{-1}\, .
\ea
Thus, the stationary solution has a Bessel-like polynomial behavior near the origin, and is regularized at large $x$ by the exponentially-decaying tail associated with the linear term in the potential. 

\begin{center}
\begin{figure}[ht]
\includegraphics[width=7cm]{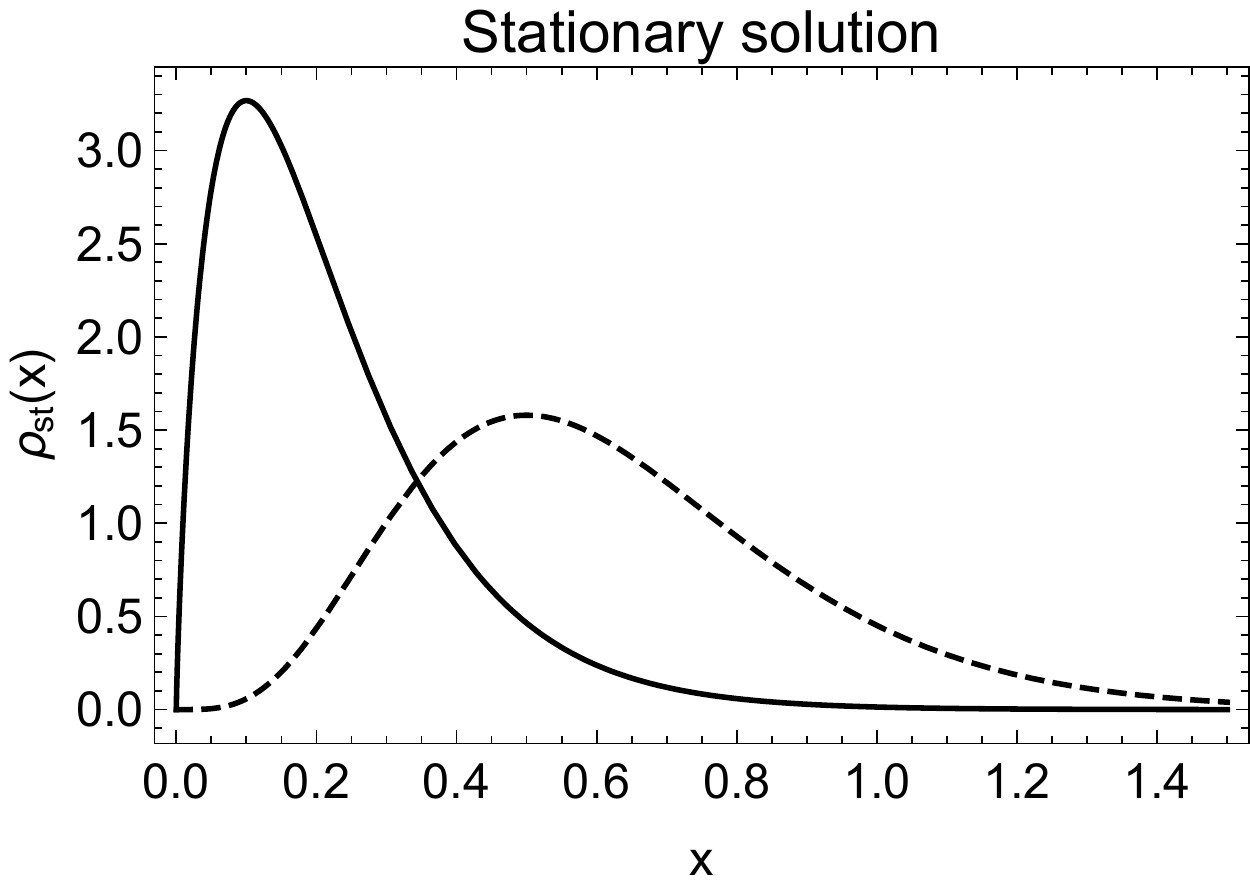}\hspace{1em}
\includegraphics[width=7cm]{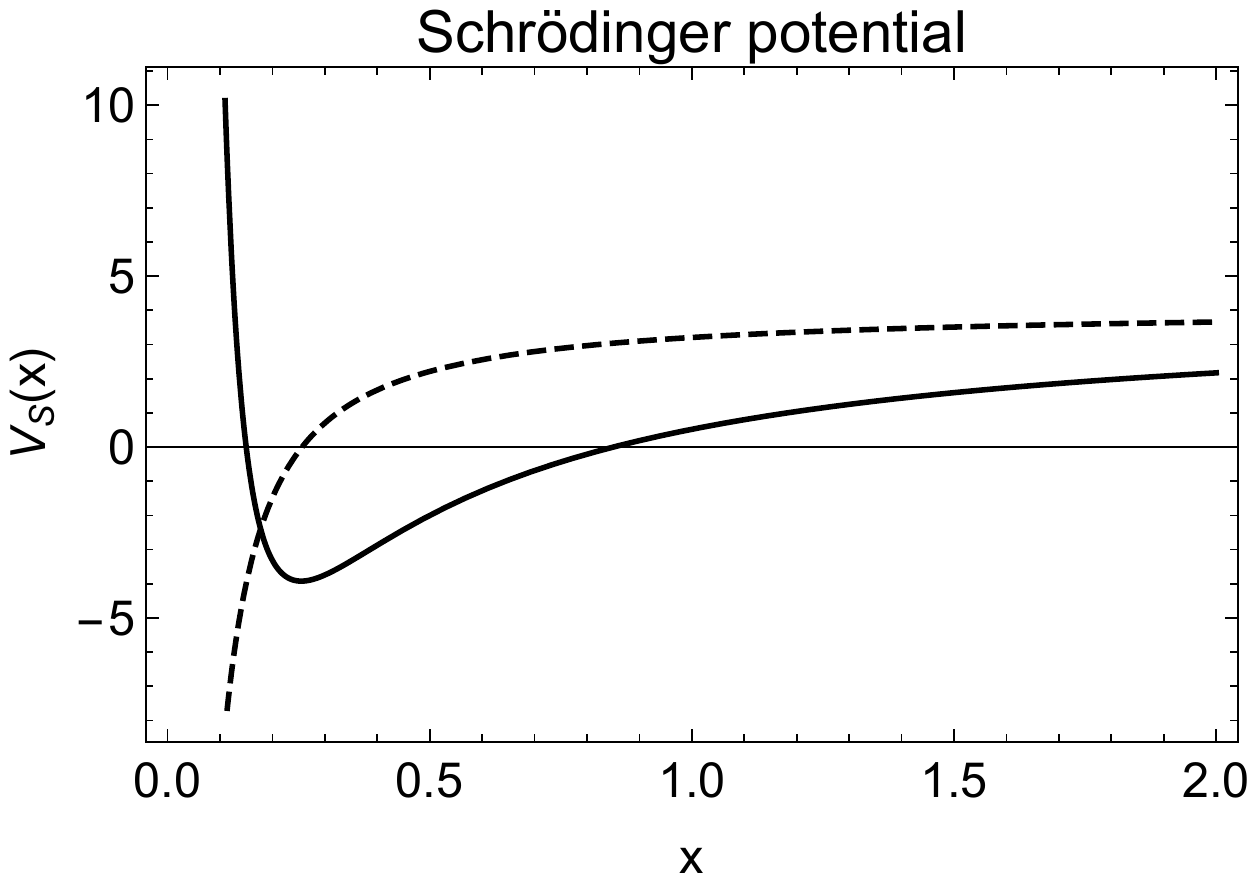}
\caption{The stationary solution (left panel) and the Schr\"odinger potential (right panel) for $b=1>\s^2$ (dashed line) and $b=0.2<\s^2$ (solid line) with $\s=0.7$.}
\label{fig:PlotCases}
\end{figure}
\end{center}

\begin{center}
\begin{figure}[ht]
\includegraphics[width=7cm]{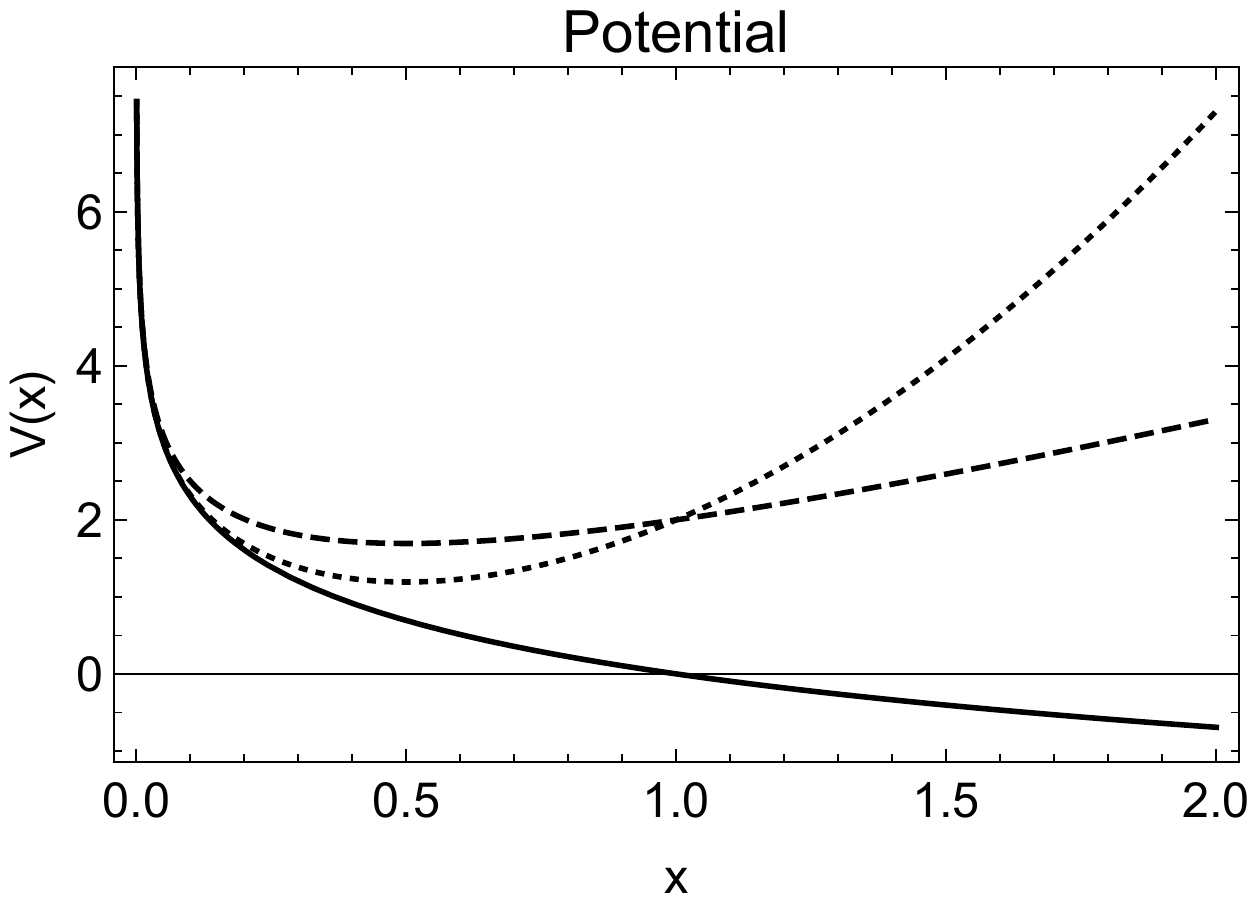}\hspace{1em}
\includegraphics[width=7cm]{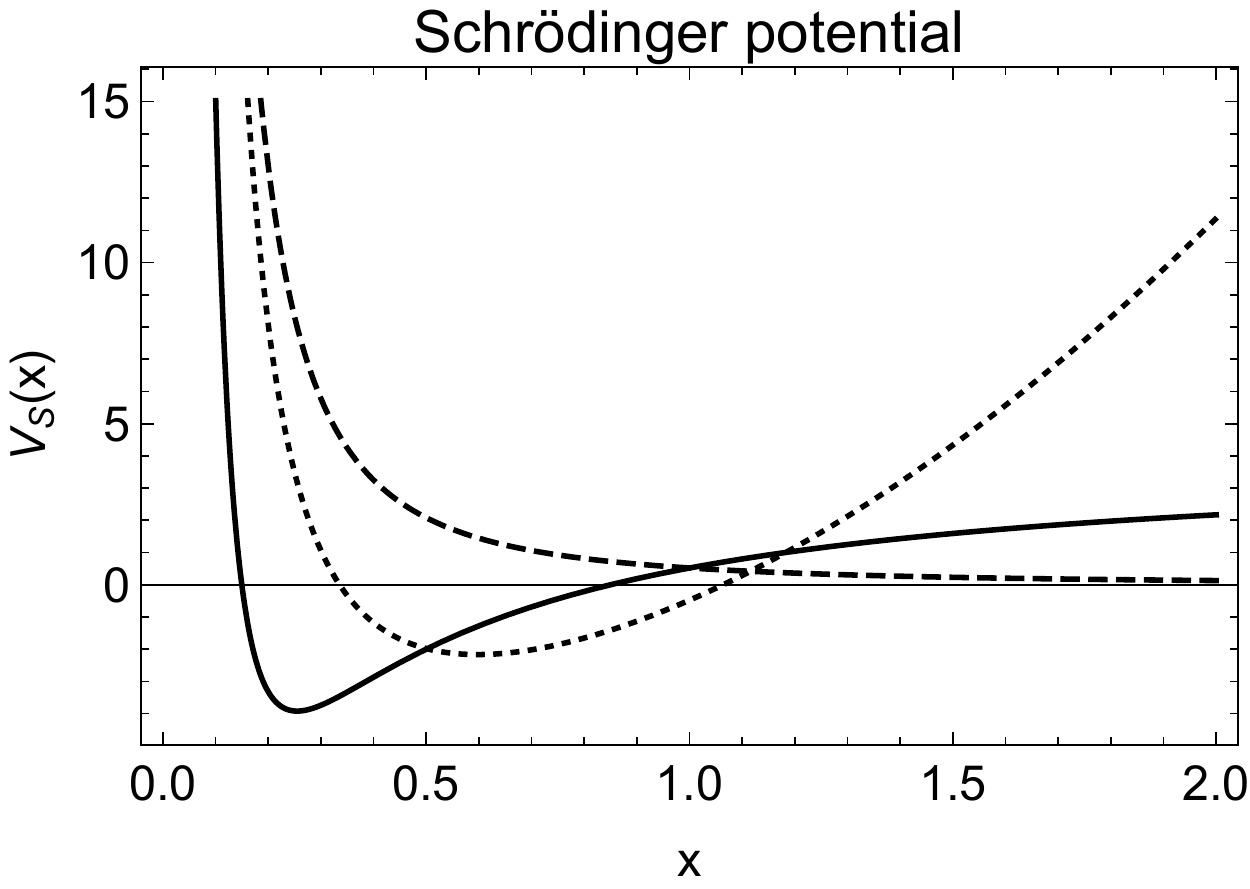}
\caption{The potentials $V(x)$ (left panel) and the Schr\"odinger potentials $\mathcal{V}_s(x)$ (right panel) of the Bessel process (dashed line), the Bessel process with constant drift (solid line) and the Rayleigh process (dotted line), for parameters $a =-2, b = 1,\s = 0.7$.}
\label{fig:PlotCases2}
\end{figure}
\end{center}

The Sturm-Liouville equation is closely related to the Schr\"odinger equation governing the radial component of the wave function of an electron with energy $E$ in an atom of atomic number $Z$, namely
\be
\p_r^2\, y(r) - \left[\frac{l(l+1)}{r^2} - \frac{2\,Z}{r}\right] \, y(r) = - 2\, E\, y(r) \, , 
\ee
where $r$ is the radial coordinate, $y(r) = r\, R(r)$, with $R(r)$ the radial component of the wavefunction, and $l$ is the angular quantum number (c.f., Eq. 2.1 of Bethe and Salpeter \citep{BSQM1957}). 
The associated Schr\"odinger potential is
\ba\label{SchrpotMINE}
\mathcal{V}_s(x) = 
\frac{b\, (b - \s^2)}{2\, \s^2}\, \frac{1}{x^2} + \frac{a\, b}{\s^2}\, \frac{1}{x} + \frac{a^2}{2\, \s^2} \, ,
\ea
which as seen in Fig. \ref{fig:PlotCases} tends asymptotically to a critical value 
\be
\Lim{x\to\infty}\mathcal{V}_s(x) \equiv \la_{cr} = \frac{a^2}{2\, \s^2} \, ,
\ee
defining the boundary between the discrete ($\la<\la_{cr}$) and the continuous ($\la>\la_{cr}$) spectra.
The Schr\"odinger potentials of the Bessel process of Eq. \eqref{FPEB} and the Rayleigh process of Eq. \eqref{Rayl}, both with $b\to -b$, are
\be
\mathcal V_s^{Bes}(x) = \frac{b\, (b-\s^2)}{2\, \s^2}\,\frac{1}{x^2}\, ,\qquad \mathcal V_s^{Ray}(x) = \frac{b\, (b-\s^2)}{2\, \s^2}\,\frac{1}{x^2} + \frac{a^2}{2\, \s^2}\, x^2 + \, a\, \left(\fr + \frac{b}{\s^2}\right) \, .
\ee
As shown in Fig.\ref{fig:PlotCases2}, whether the potentials and the Schr\"odinger potentials decay or diverge asymptotically leads to either continuum or discrete states.  A peculiarity of the Bessel process is that, due to a scaling symmetry of the potential, its eigenfunctions are not oscillatory as we might expect, but instead decay at infinity \cite{ESGR2005}.
Now we describe the technique of solving the Sch\"odinger-like equation for the Bessel process with constant drift. 
We note that when reparameterizing the equation in terms of Whittaker functions (see Appendix \ref{AppendixWhittaker}), it describes an equivalent process with multiplicative noise \cite{SBMS1979}. The generic solution $u_\la(x)$ of the Schr\"odinger equation reads
\be\label{eigval}
u_\la(c) = u^{(1)}_\la(x) + u^{(2)}_\la(x) = \text{C}_1\, M_{\left\{-\frac{a\, b}{\s^2\, \zeta_\la}, \fr - \frac{b}{\s^2}\right\}}\left(\frac{2\, x\, \zeta_\la}{\s^2}\right) + \text{C}_2\, W_{\left\{-\frac{a\, b}{\s^2\, \zeta_\la}, \fr - \frac{b}{\s^2}\right\}}\left(\frac{2\, x\, \zeta_\la}{\s^2}\right)\, ,
\ee
where $M$ and $W$ are the Whittaker functions of the first- and second-kind, $\text{C}_1$ and $\text{C}_2$ are the integration constants, and 
\be\label{zetapar}
\zeta_\lambda = \sqrt{a^2 - 2\, \lambda\, \s^2}\, .
\ee
By using the relations 
\be\label{hyperf1}
M_{\{\alpha,\beta\}}(x) = \text{e}^{- \frac{x}{2}}\, x^{\beta + \fr}\, M\left(\beta - \alpha + \fr, 1 +  2\, \beta; x\right) \, ,
\ee
and
\be\label{hyperf2}
W_{\{\alpha,\beta\}}(x) = \text{e}^{- \frac{x}{2}}\, x^{\beta + \fr}\, U\left(\beta - \alpha + \fr, 1 +  2\, \beta; x\right)\, ,
\ee
we can rewrite the solution as
\ba\label{solutions}
u_{\la}(x) = \text{e}^{- \frac{\zeta_\lambda}{\s^2}\,x}\, \left(\frac{2\, x\, \zeta_\lambda}{\s^2}\right)^{1 - \frac{b}{\s^2}}\left\{\hspace{1em}\text{C}_1\, M\left(1 - \frac{b}{\s^2} + \frac{a\, b}{\s^2\, \zeta_\lambda}, 2 - \frac{2\, b}{\s^2}; \frac{2\, x\, \zeta_\lambda}{\s^2}\right) \hspace{2.5em}\right. \nonumber\\ 
\left. + \, \text{C}_2\, U\left(1 - \frac{b}{\s^2} + \frac{a\, b}{\s^2\, \zeta_\lambda}, 2 - \frac{2\, b}{\s^2}; \frac{2\, x\, \zeta_\lambda}{\s^2}\right) \hspace{1em}\right\} \, ,
\ea
where $M$ and $U$ are the hypergeometric confluent functions of the first- and second-kind respectively, or the Kummer ($M$) and Tricomi ($U$) functions.  Now, by using the symmetry property
\be\label{symm}
W_{\{\alpha,\beta\}}(x) = W_{\{\alpha,-\beta\}}(x) \, ,
\ee
we can write the $W$-functions as
\be
\label{Wexp2}
u^{(2)}_\la(x) = \text{C}_2\, W_{\left\{-\frac{a\, b}{\s^2\, \zeta_\la}, \fr - \frac{b}{\s^2}\right\}}\left(\frac{2\, x\, \zeta_\la}{\s^2}\right)\, = \text{C}_2\,  \text{e}^{-\frac{\zeta_\la}{\s^2}\, x}\, \left(\frac{2\, x\, \zeta_\la}{\s^2}\right)^{\frac{b}{\s^2}}\, U\left(\frac{b}{\s^2}+\frac{a\, b}{\s^2\, \zeta_\la},\frac{2\, b}{\s^2};\frac{2\, x\, \zeta_\la}{\s^2}\right) \, .
\ee
%
%
\subsection{Discrete spectrum}
By taking the limit $\la\to0$ of Eqs. \eqref{solutions} and \eqref{Wexp2}, and considering that 
\be
\Lim{\la\to0} \zeta_\la = \sqrt{a^2} = - a\, ,
\ee
we find that only the $W$-function provides the stationary solution,
\be
W_{\left\{-\frac{a\, b}{\sqrt{a^2}\, \s^2}, \fr - \frac{b}{\s^2}\right\}}\left(\frac{2\, \sqrt{a^2}\, x}{\s^2}\right)
= \text{e}^{\frac{a\, x}{\s^2}}\, x^{\frac{b}{\s^2}} \, \left\{ \left(- \frac{2\,a}{\s^2}\, x\right)^{\frac{b}{\s^2}}\, U\left(0, \frac{b}{\s^2}; -\frac{2\, a\, x}{\s^2} \right) \right\} \, , 
\ee
where $U(0,\beta;x) = 1$.  Hence, from here forward we set $\text{C}_1 = 0$. 

Upon imposition of the boundary conditions, $\psi_\la(x=0) = \psi_\la(x=\infty) = 0$, we note that the condition at infinity is always fulfilled, whereas that at the origin is only satisfied when the first argument is a negative integer, thereby defining the discrete set of eigenvalues as
\be\label{nva}
\frac{b}{\s^2} + \frac{a\, b}{\s^2\, \zeta_\lambda} = - n\, ,
\ee
with $n$ a non-negative integer so that
\be\label{eigenb}
\lambda_n = \frac{(a\, n\, \s)^2 + 2\, a^2\, b\, n}{2\, (b + n\, \s^2)^2}\, .
\ee
For negative values of its first argument, the Tricomi function reduces to the unassociated Laguerre polynomial viz., 
\be\label{Leg}
U(n,k, \gamma\, x) = \frac{(-n)!}{(-1)^{-n}}\, L_{-n}^{k-1}(\gamma\, x) \,.
\ee
Substituting Eq. \eqref{eigenb} into the expression \eqref{zetapar} of $\zeta_\la$ gives
\be
\zeta_\lambda = \sqrt{a^2 - 2\, \lambda\, \s^2} = \sqrt{\frac{(a\, b)^2}{(b + n\, \s^2)^2}} = -\frac{a\, b}{(b + n\, \s^2)}\, ,
\ee
and substituting this into $u^{(2)}_\la(x)$, Eq. \eqref{Wexp2}, and using \eqref{Leg} leads to the expression
\ba\label{eigun}
u^{(2)}_\la(x) = \text{C}_2\, \left(\xi_n\, x\right)^{\frac{b}{\s^2}}\, \text{e}^{- \frac{\xi_n}{2}\, x}\, U\left(-n, \frac{2\, b}{\s^2}, \xi_n\, x\right)  = \text{C}_2\, \left(\xi_n\, x\right)^{\frac{b}{\s^2}}\, \text{e}^{- \frac{\xi_n}{2}\, x}\,\frac{n!}{(-1)^n}\, L_n^{-1+\frac{2\, b}{\s^2}}(\xi_n\, x) \, , \nonumber \\
\ea
in which
\be
\xi_n = - \frac{2\, a\, b}{\s^2\, (b + n\, \s^2)} \, .
\ee
The normalization $\text{C}_2$ is determined by the integral
\be
\int_0^\infty\, dx\, \left[\text{e}^{- \frac{c\, x}{2}}\, (c\, x)^{\frac{a+1}{2}}\, L_n^a(c\, x)\right]^2 = \frac{\Gamma(n+a+1)\, (2\, n + a +1)}{c\, n!}\, ,
\ee
which gives
\be
\text{C}_2 = \left(\frac{\xi_n\, n!}{\Gamma\left(n + \frac{2\, b}{\s^2}\right)\, \left(2\,n + \frac{2\,b}{\s^2}\right)}\right)^\fr\, .
\ee
Finally, the eigenfunctions in the discrete spectrum region are
\be\label{discspec}
\psi_n^{dis}(x) = 
\left(\frac{\xi_n\, n!}{\Gamma\left(n + \frac{2\, b}{\s^2}\right)\, \left(2\,n + \frac{2\,b}{\s^2}\right)}\right)^{\fr} \, \left(\xi_n\, x\right)^{\frac{b}{\s^2}}\, \text{e}^{-\frac{\xi_n}{2}\, x}\, L_n^{-1 +\frac{2\, b}{\s^2}}\left(\xi_n\, x\right) \, ,
\ee
which form an infinite set (hence $N=\infty$ in the spectral representation of Eq. \eqref{mix}) that are orthonormalized in the sense Eq. \eqref{normdisc}. 
However, the complete set is the sum of the eigenfunctions of the discrete {\em and} continuous spectrum\footnote{Completeness can also be inferred by reparameterizing the solution with $\tilde x = \xi_n\, x$ and $\gamma=-1+(2\, b)/\s^2$ obtaining $\psi_n(\tilde x) = \mathcal C\, \omega(\tilde x)\, L_n^{\gamma}(\tilde x)$, with $\omega(\tilde x) =\text{e}^{-\tilde x/2}\, \tilde x^{(\gamma+1)/2}$ and $\mathcal C$ a constant. The weight function of a complete basis of Laguerre polynomials is instead $\omega(\tilde x) =\text{e}^{-\tilde x/2}\, \tilde x^{\gamma/2}$.  In other words, a complete orthonormal basis of Laguerre polynomials is found by letting $\left(\xi_n\, x\right)^{\frac{b}{\s^2}} \to \left(\xi_n\, x\right)^{\frac{b}{\s^2}-\fr}$ in Eq. \eqref{discspec}.}. 
\begin{center}
\begin{figure}[ht]
\includegraphics[width=7cm]{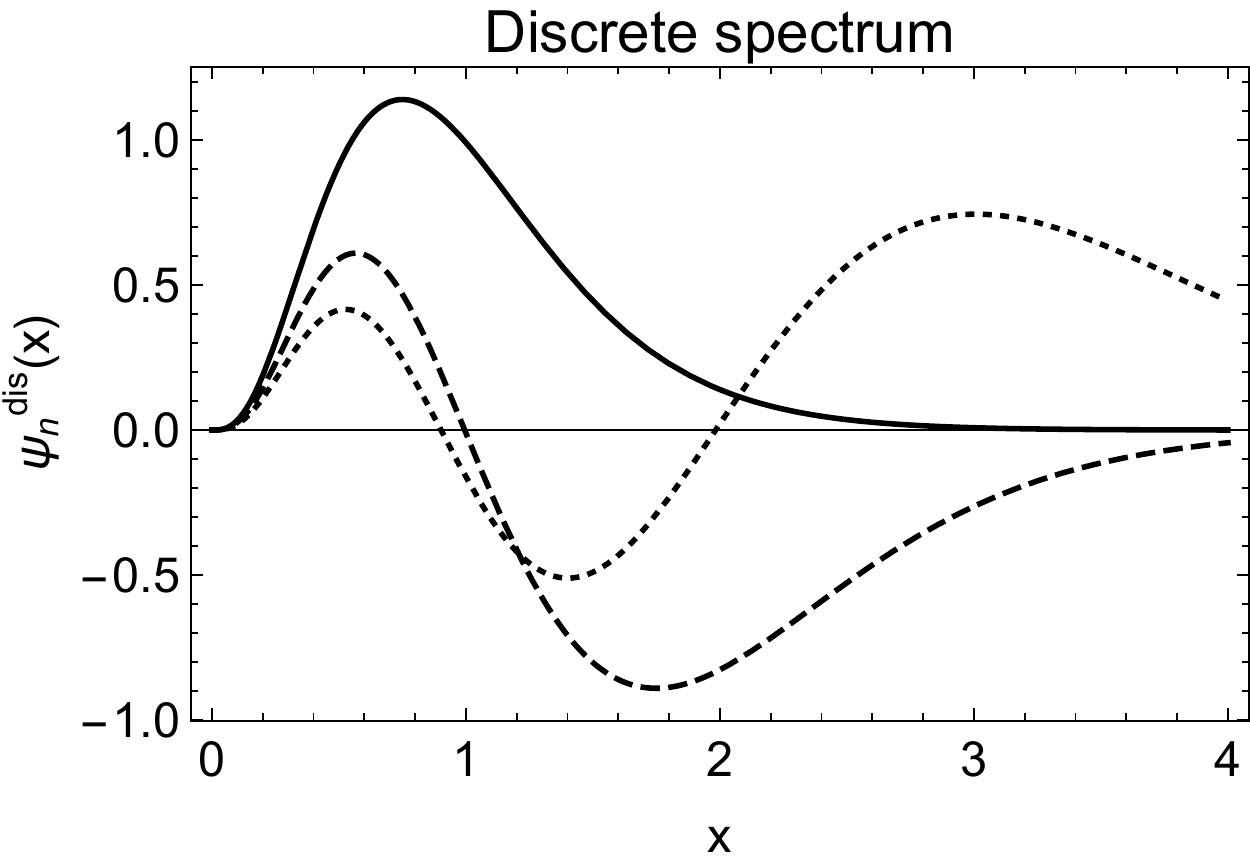}\hspace{1em}
\includegraphics[width=7cm]{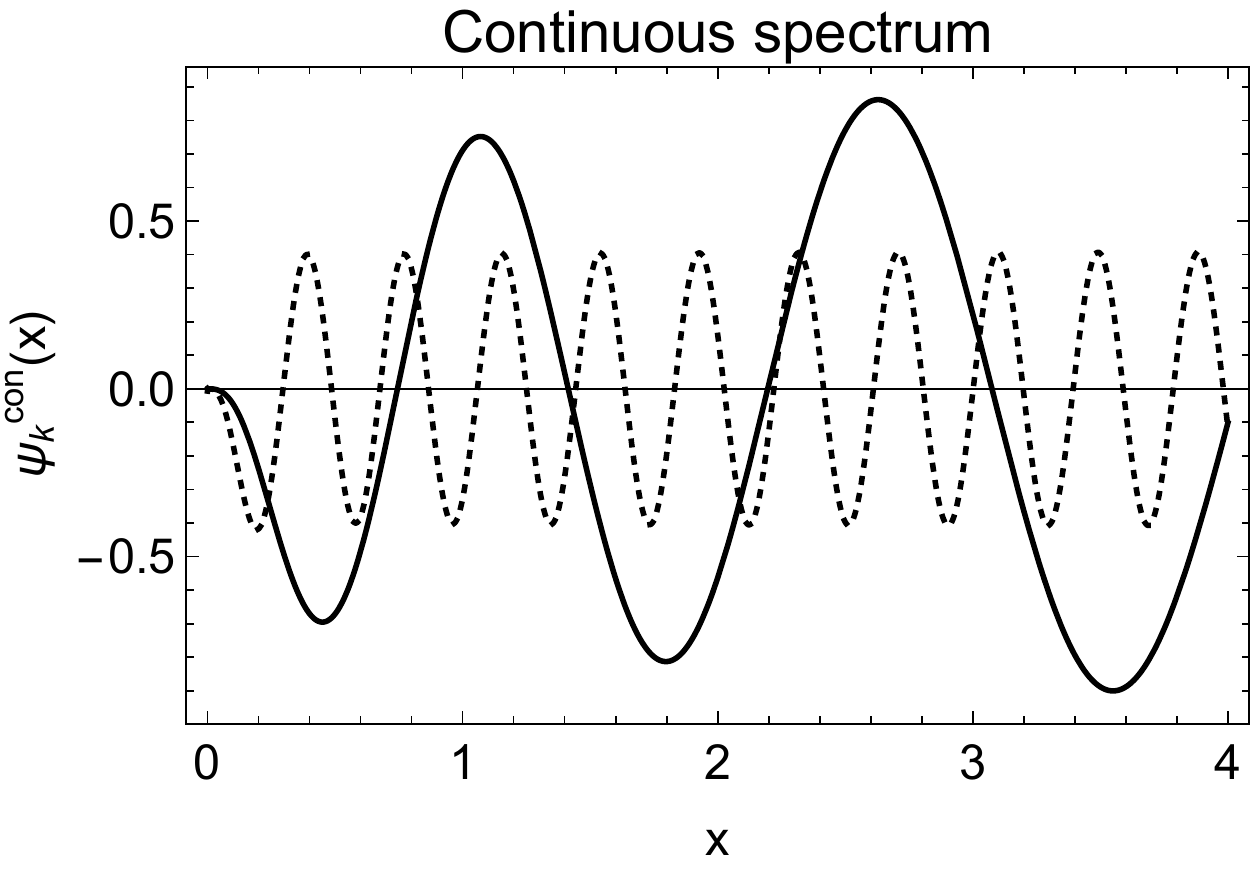}
\caption{Eigenfunctions for $a=-2, b=1.5$. and $\s=0.7$. Discrete spectrum in the left panel for $n=0$ (solid line), $n=1$ (dashed line), and $n=2$ (dotted line). Continuous spectrum in the right panel for $k=0.5$ (solid line) and $k=30$ (dotted line). }
\label{fig:PlotEigenf}
\end{figure}
\end{center}
%
%
\subsection{Continuous spectrum}
In the continuous regime $\la > \la_{cr} = a^2/(2\, \s^2)$ we are only concerned with the boundary condition at the origin; $\psi_\la^{con}(x=0)=0$. Note that for $\la > \la_{cr}$ the parameter $\zeta_\la$ defined in Eq. \eqref{zetapar} takes on complex values. Therefore, we take only the real part of the complex eigenfunctions and parameterize the eigenvalues as
\be
\lambda = \lambda_{cr} + k\, ,
\ee
with 
$k\geq0$. To impose $\psi_\la^{con}(x=0)=0$ we study the behavior of the general solution at small $x$. Both solutions diverge at $x=0$ for $b > \s^2$ and vanish for $b < \s^2$. In the former case we have
\ba\label{compwhi}
\lim_{x \underset{>}{\to} 0}~M_{\left\{\frac{i\, a\, b}{\sqrt{2~k}\, \s^3},\frac{1}{2}-\frac{b}{\s^2}\right\}}\left(\frac{i\, 2\, \sqrt{2~k}}{\s}\, x\right) \hspace{1em} = \hspace{1em}  2^{\frac{3}{2}-\frac{3\, b}{2\, \s^2}}\, \left(\frac{i}{\s}\right)^{1-\frac{b}{\s^2}}\, k^{\frac{1}{2}-\frac{b}{2\, \s^2}} x^{1-\frac{b}{\s^2}}\, , \hspace{3.3em}\\
\lim_{x \underset{>}{\to} 0}~W_{\left\{\frac{i\, a\, b}{\sqrt{2~k}\, \s^3},\frac{1}{2}-\frac{b}{\s^2}\right\}}\left(\frac{i\, 2\, \sqrt{2~k}}{\s}\, x\right) \hspace{1em} = \hspace{1em} 2^{\frac{3}{2}-\frac{3 b}{2 \sigma ^2}} \left(\frac{i}{\s}\right)^{1-\frac{b}{\s^2}}\, k^{\frac{1}{2}-\frac{b}{2 \sigma^2}}  x^{1-\frac{b}{\sigma^2}} \, \frac{\Gamma \left(\frac{2 b}{\sigma ^2}-1\right)}{\Gamma \left(\frac{b}{\s^2} - \frac{i\, a\, b}{\sqrt{2~k}\, \s^3}\right)}\nonumber.
\ea
Thus, by taking a linear combination of the $W$- and $M$-functions we define a modified Whittaker function, $K$, in which the divergencies at the origin cancel each other; 
\be\label{modwhit}
K_{\{\alpha, \beta\}}(\gamma\, x) = M_{\{\alpha,\beta\}}(x) - \frac{\Gamma \left(\frac{b}{\s^2} - \frac{i\, a\, b}{\sqrt{2}\, \sqrt{k}\, \s^3}\right)}{\Gamma \left(\frac{2\, b}{\s^2}-1\right)}\, W_{\{\alpha, \beta\}}(x) \, ,
\ee
and thus $K$ vanishes at the origin, where $\alpha$, $\beta$ and $\gamma$ are the arguments of the functions in Eq. \eqref{compwhi}. 

Finally, Eq. \eqref{normhalf1} gives the normalization of the eigenfunctions, which are
\be\label{eigfcont1}
\psi_k^{con}(x) = \mathcal{N}\, \Re \left[K_{\left\{\frac{i\, a\, b}{\sqrt{2}\, \sqrt{k}\, \s^3},\frac{1}{2}-\frac{b}{\s^2}\right\}}\left(\frac{i\, 2\, \sqrt{2}\, \sqrt{k}}{\s}\, x\right)\right]\, , \qquad
\mathcal{N} = \frac{2^{\frac{1}{4}}}{\sqrt{\s\, \pi}\, k^{\frac{1}{4}}\,\mathcal Y(k)}\, .
\ee
where $\Re$ is the real part and $\mathcal Y(k)$ is the amplitude of asymptotic oscillations. Note that Eq. \eqref{eigfcont1} also holds when $b<\s^2$. The evaluation and precision of the normalization coefficient are described in Appendix \ref{AppendixNormW}.
%
%
\subsection{Numerical comparison}
\begin{center}
\begin{figure}[ht]
\includegraphics[width=7cm]{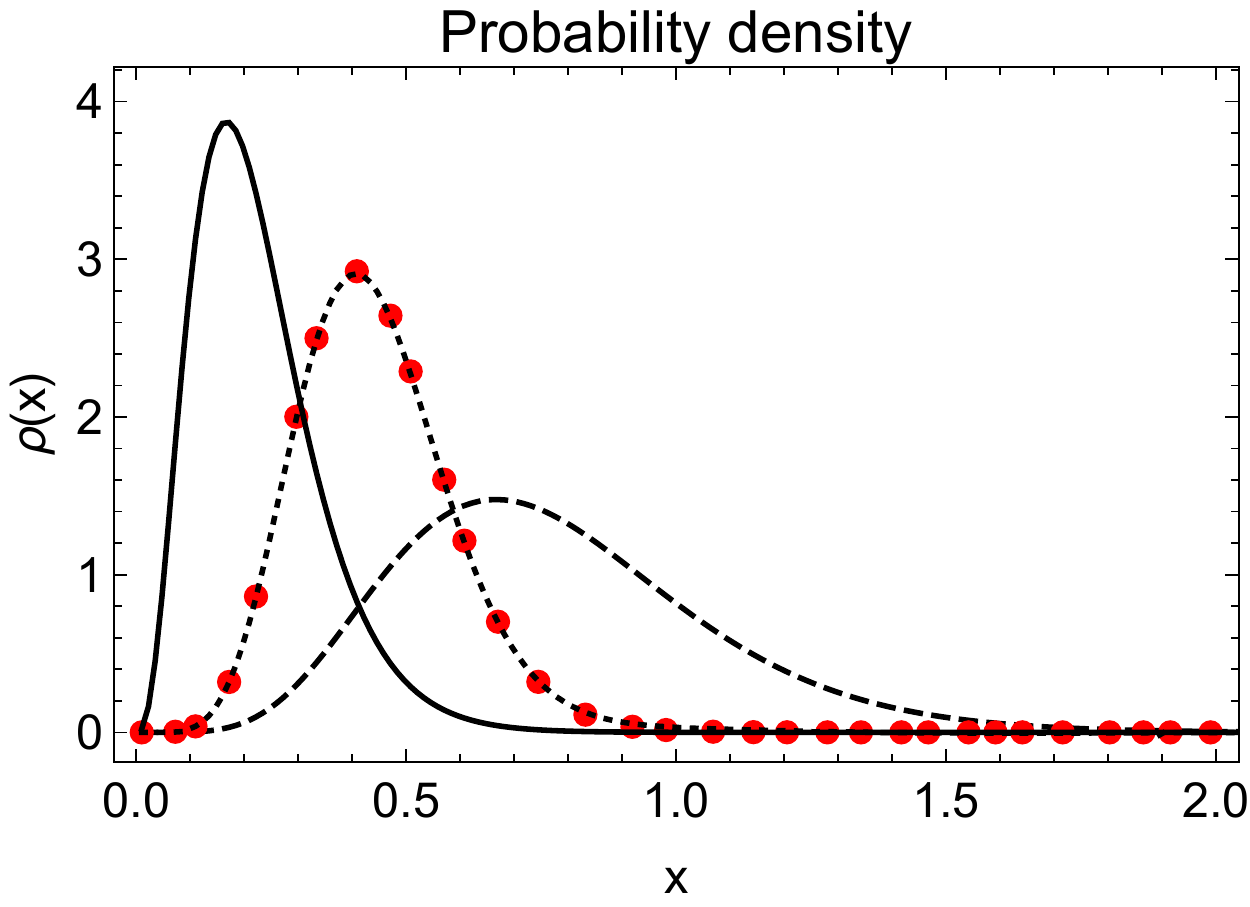}\hspace{1em}
\includegraphics[width=7.2cm]{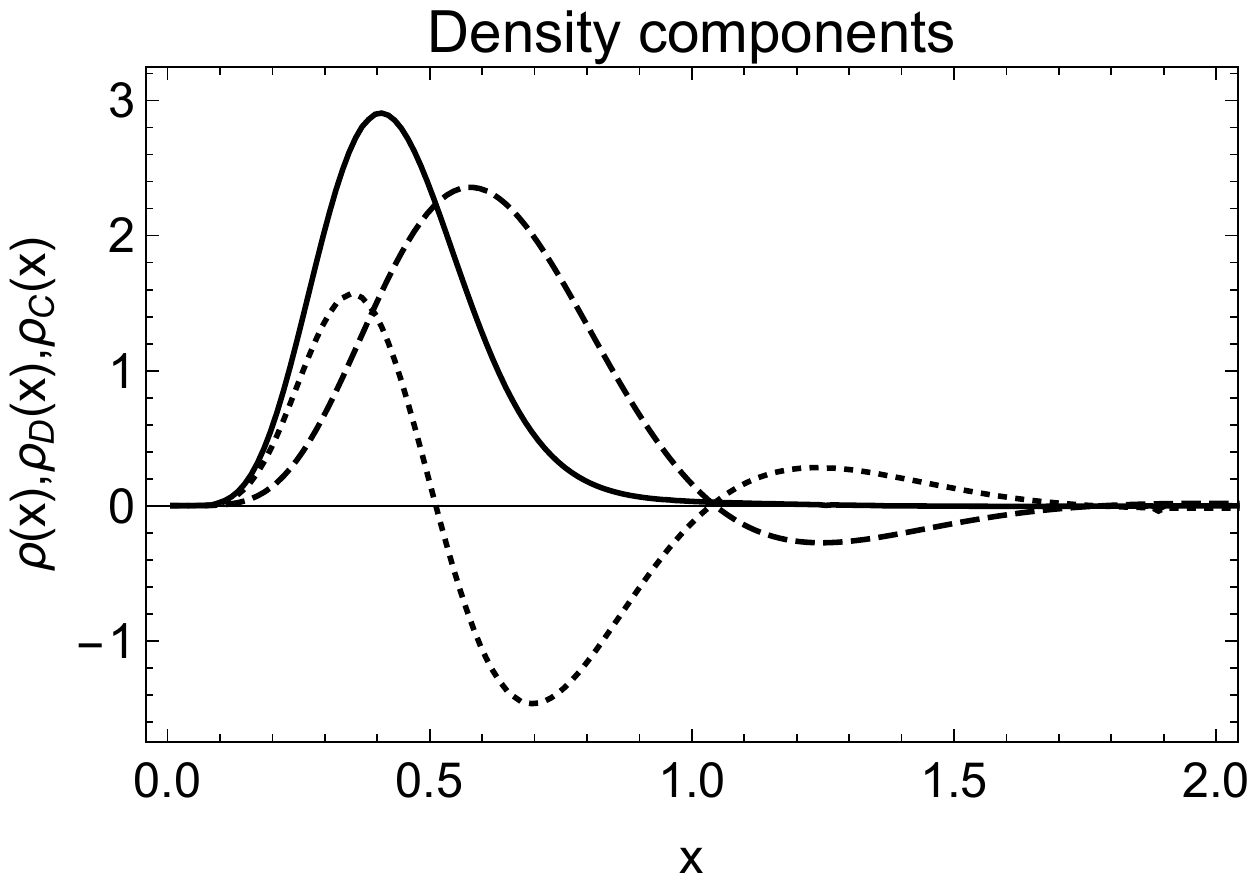}
\caption{Probability density for $a=-3, b=2$ and $\s=0.8$. Left panel: initial condition ($T=0$, solid line), analytic solution at $T=0.06$ (dotted line), Crank-Nicholson solution (red dots in the color version), stationary solution ($T=\infty$, dashed line). Right panel: full analytic solution at $T=0.06$ (solid line), sum over the discrete spectrum (dashed line), integral over the continuous spectrum (dotted line). Both the analytical expression and the Crank-Nicholson algorithm were written in Wolfram Mathematica 10. The plot was evaluated using 20 eigenfunctions in the discrete spectrum and an AccuracyGoal equal to 3 for the continuum. Because our solution is written in terms of a complicated asymptotic linear combination, Mathematica took approximately five times longer in absolute time to evaluate the integral over $k$ relative to the solution of Linetsky \cite{LINE2004}.}
\label{fig:PlotComparison1}
\end{figure}
\end{center}

The spectral expansion of the transition density is
\ba
\rho(x,t|x_0,t_0) = \sqrt{\frac{\rho_{st}(x)}{\rho_{st}(x_0)}} \left\{ 
\sum_{n=0}^{\infty}\, \text{e}^{- \la_n\, (t-t_0)}\, \psi_{n}^{dis}(x_0)\, \psi_{n}^{dis}(x)  \nonumber  \right. \hspace{8em} \\ \left.
+ \text{e}^{- \la_{cr}\, (t-t_0)}\, \int_{0}^{\infty}\, dk\, \text{e}^{- k\, (t-t_0)}\, \psi_k^{con}(x_0)\, \psi_k^{con}(x) 
\right\} \, ,
\ea
where $\rho_{st}(x)$, $\psi_{n}^{dis}(x)$ and $\psi_k^{con}(x)$ are Eqs. \eqref{rhos}, \eqref{discspec} and \eqref{eigfcont1} respectively. Note that our expression is equivalent to that provided by Linetsky (see Eqs. 6 and 8 of \cite{LINE2004}). However, while the form of the eigenfunctions of the discrete spectrum coincides with his, the continuum is written in terms of different Whittaker functions; complex $M$-functions in Linetsky's work and a linear combination of $W$ and $M$-functions in our work. Of course the two formulations must be equivalent. In the left panel of Fig. \ref{fig:PlotComparison1} we compare our solution for the probability density to a numerical solution evaluated using the Crank-Nicholson method with a time step of $\Delta t = 10^{-5}$ and an initial condition of $\rho(x_0,t_0) \propto x_0^\beta\, \text{e}^{\, \alpha\, x_0}$, where $\alpha= -16$ and $\beta=3$.  
%
%
\vspace{-0.5in}
\section{\label{sec:CL}Conclusion}
The technique described in this paper is motivated by a stochastic model in climate dynamics, which treats the seasonal evolution of the Arctic sea ice thickness.  The summer is modeled by a Brownian motion process with a negative constant drift and the winter by a Bessel process with a constant negative drift.  We have solved the associated Fokker-Planck equations whose spectral expansions exhibit a continuum or a mixed (discrete and continuum) spectrum of eigenvalues. Relying upon early work in quantum mechanics, we posited that the normalization of oscillating eigenfunctions in the continuous spectrum depends solely upon the amplitude of the asymptotic oscillations, which led to a general formula for the solution of the one-dimensional autonomous Fokker-Planck equation with additive noise. We have employed this technique to solve the problem of Brownian motion with constant drift and natural boundary conditions at infinity and an absorbing boundary at the origin. Such well-trodden ground served as a good test bed for the approach, which we then demonstrated by solving 
the more complicated equation arising from the Bessel process with negative constant drift. Our solution coincides with that of Linetsky who employed different methods \cite{VALI2004,LINE2004}.  However, the simplicity of our method of normalizing the functions in the continuum comes at the price of a lower computational efficiency.  Therefore, depending on the specifics of the application, one may consider both methods on equal footing.  Our approach can be used to examine the fully time dependent climatological evolution of the system by matching the two processes continuously, which is the subject of a future study. 
\vspace{-0.2in}
\begin{acknowledgments}
We thank S. Agarwal, F. Mancarella, M. Testa and S. Toppaladoddi for useful discussions. W.M. acknowledges a NASA Graduate Research Fellowship and a Herchel-Smith postdoctoral fellowship for support. W.M. acknowledges a Herchel-Smith postdoctoral fellowship for support. J.S.W. acknowledges NASA Grant NNH13ZDA001N-CRYO, Swedish Research Council grant no. 638-2013-9243, and a Royal Society Wolfson Research Merit Award for support. All authors acknowledge the 2015
Geophysical Fluid Dynamics Summer Study Program at the Woods Hole Oceanographic Institution, which is
supported by the National Science Foundation and the Office of Naval Research under OCE-1332750.
\end{acknowledgments}

\numberwithin{equation}{section}
\begin{appendices}
%
\section{\label{AppendixFues}Determination of prefactor $\Delta_n k$}
We start by considering the normalization relation for eigendifferentials,
\be
\frac{1}{\Delta_n k}\, \int_\Omega \, dx \, \Psi_n(x) \, \Psi_m(x) = \delta_{nm} \, .
\ee
We then use the definition in Eq. \eqref{eigendiff} to rewrite the eigendifferentials in terms of eigenfunctions. Performing the spatial integral first, we obtain 
\be
\frac{1}{\Delta_n k}\, \int_{\Delta_n k}\, dk_1\, \int_{\Delta_m k}\, dk_2 \int_\Omega \, dx\, \psi_{k_1}^{con}(x)\, \, \psi_{k_2}^{con}(x) = 1 \, .
\ee
By using Eq. \eqref{dirac} we arrive at
\be
\frac{1}{\Delta_n k}\, \int_{\Delta_n k}\, dk_1\, \int_{\Delta_m k}\, dk_2 \,\,  \delta(k_1-k_2)\, = 1 \, .
\ee
When evaluating the integral in $k_2$ for $m\neq n$, since $k_1$ lies outside of the range of integration of $k_2$, the delta function gives zero, thereby reproducing the behavior of the Kronecker delta. If instead $m=n$, the integration gives
\be
\frac{1}{\Delta_n k}\, \int_{\Delta_n k}\, dk_1\, = 1 \, .
\ee
%
\section{\label{AppendixWhittaker}Parameterization into the Whittaker equation}
Here we parameterize the Sturm-Liouville equation with the potential in Eq. \eqref{SchrpotMINE} into the Whittaker equation, 
\be\label{whit1}
\left\{d_x^2 - \left(\frac{1}{4} - \frac{k}{x} - \frac{\frac{1}{4} - n^2}{x^2}\right)\right\}\, f(x) = 0\, ,
\ee
which has the general solution
\be
f(x) = \text{C}_1\, W_{\{k,n\}}(x) + \text{C}_2\, M_{\{k,n\}}(x) \, ,
\ee
where $\text{C}_1$ and $\text{C}_2$ are integration constants, and  $M_{\{k,n\}}(x)$ and $W_{\{k,n\}}(x)$ are Whittaker functions of the first- and second-kind respectively, and are  related to the confluent hypergeometric functions of the first- and second-kind via Eqs. \eqref{hyperf1} and \eqref{hyperf2}. 
The confluent functions are related to the unassociated Laguerre polynomials through
\be\label{Lagr}
L_n^k(x) = \binom{n+k}{n}\, M(-n, k+1; x) = \frac{(-1)^n}{n!}\, U(-n, k+1; x)\, ,
\ee
where $n$ is a positive natural number and $k$ and $x$ are real numbers.  Finally we note that the $W$-function satisfies the identify $W_{\{\alpha,\beta\}}(x) = W_{\{\alpha,-\beta\}}(x)$.

We rescale Eq. \eqref{whit1}  with $x \to \alpha\, x$, so that the equation and its solution are now
\ba\label{whit2}
\left\{\frac{1}{\alpha^2}\, d_x^2 - \left(\frac{1}{4} - \frac{k}{\alpha\, x} - \frac{\frac{1}{4} - n^2}{(\alpha\, x)^2}\right)\right\}\, f(x) = 0\, ,\\[0.5em]
f(x) = c_1\, W_{\{k,n\}}(\alpha\, x) + c_2\, M_{\{k,n\}}(\alpha\, x) \, .
\ea
Multiplying  both sides with $\beta$, setting $\alpha=\sqrt{2\, \beta}/\sigma$ and rescaling $k \to k/\sqrt{\beta}$ we find
\ba\label{whit3}
\left\{\frac{\sigma^2}{2}\, d_x^2 - \left(\frac{\beta}{4} - \frac{k\, \sigma}{\sqrt{2}\, x} - \frac{\left(\frac{1}{4} - n^2\right)\, \sigma^2}{2\, x^2}\right)\right\}\, f(x) = 0\, , \\[0.5em]
f(x) = c_1\, W_{\{\frac{k}{\sqrt{\beta}},n\}}\left(\frac{\sqrt{2\, \beta}}{\sigma}\, x\right) + c_2\, W_{\{\frac{k}{\sqrt{\beta}},n\}}\left(\frac{\sqrt{2\, \beta}}{\sigma}\, x\right) \, .
\ea
Finally, we obtain Eq. \eqref{SchrpotMINE} by letting
\be
\beta=4\,\left(\frac{a^2}{2\, \s^2} - \lambda\right)\, , \qquad k = - \frac{\sqrt{2}\, a\, b}{\s^3}\, , \qquad n = \sqrt{\frac{b\, (b-\s^2)}{\s^4} + \frac{1}{4}} = \fr - \frac{b}{\s^2}\, . 
\ee
%
\section{\label{AppendixNormW}Normalization of the modified \\ Whittaker function, $K$.}
\begin{center}
\begin{figure}[ht]
\includegraphics[width=10cm]{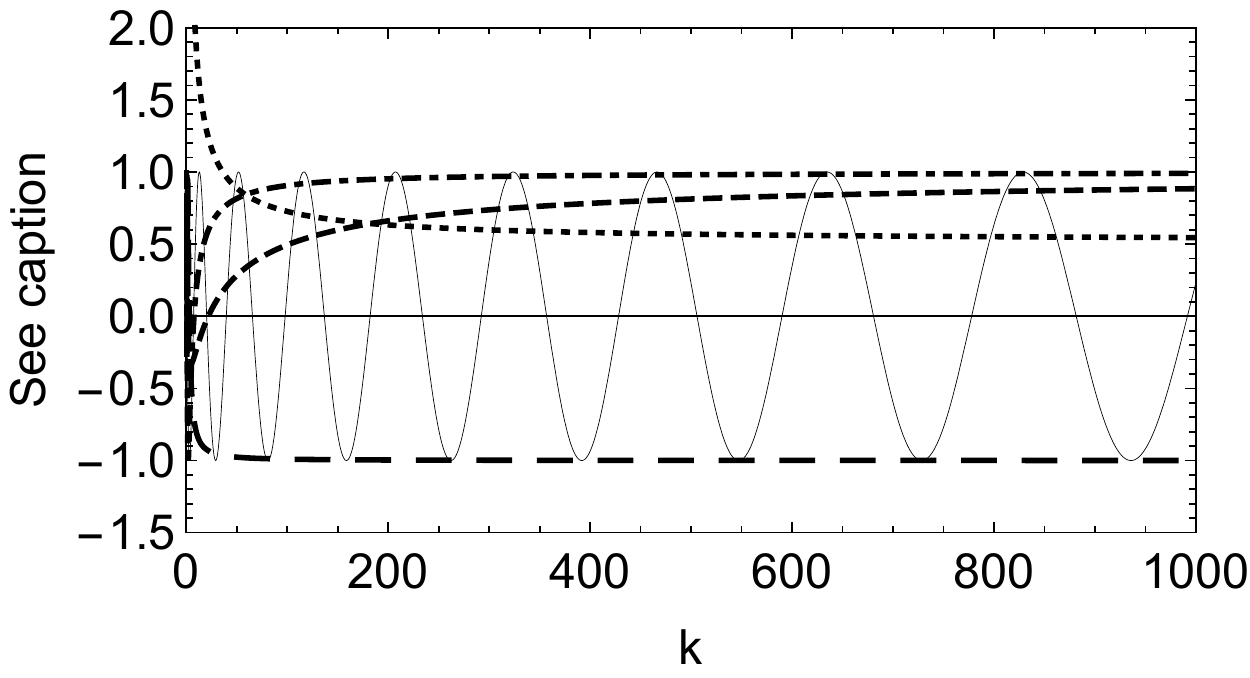}
\caption{Asymptotic amplitude and phases of the $K$-function for $a=-3, b=1.6$, $\s=0.9$ and $x=1$. The functions plotted are the amplitude $\mathcal Y(k)$ (dotted line), the complex exponential with phase $\phi_1 \propto \sqrt{k}\,x$ (solid line), $\phi_2 \propto \log(\sqrt{k}\,x)/\sqrt{k}$ (dashed line), $\phi_3 \propto 1/\sqrt{k}$ (dot-dashed line), and $\phi_4 = \mathcal W(k)$ (long dashed line). }
\label{fig:PlotPhases}
\end{figure}
\end{center}

Here we evaluate the asymptotic amplitude of the oscillating solutions in the continuous spectrum, and numerically test the accuracy of Eq. \eqref{normhalf1}.
We start by taking the asymptotic behavior of Eq. \eqref{modwhit} in a polar representation viz.,%
\ba
&&\lim_{x {\to} \infty}~K_{\left\{\frac{i\, a\, b}{\sqrt{2~k}\, \s^3},\frac{1}{2}-\frac{b}{\s^2}\right\}}\left(\frac{i\, 2\, \sqrt{2~k}}{\s}\, x\right) =  \nonumber \\
&&\text{e}^{i\, \{\phi_1(k,x) + \phi_2(k,x) + \phi_3(k)\}}\, \{ \mathcal A(k)\, \text{e}^{\frac{i\, \pi\,  b}{\s^2}} + \mathcal B(k)\, \text{e}^{i\, 0} + \mathcal C(k)\, \text{e}^{i\,  \left(\frac{\pi\,  b}{\s^2} - \frac{\pi}{2}\right)} + \mathcal D(k)\, \text{e}^{- i\, \frac{\pi}{2}}\}\, ,
\ea
where
\be
\phi_1(k,x) = \frac{\sqrt{2~k}\, x}{\s}\, ,\qquad \phi_2(k,x) = - \frac{a\, b\, \log \left(\frac{\sqrt{k}\, x}{\s}\right)}{\sqrt{2~k}\, \s^3} \, , \qquad
\phi_3(k) = - \frac{a\, b\, \log(8)}{2\, \sqrt{2~k}\, \s^3}\, ,
\ee
with
\ba\label{asympKoeff}
\mathcal A(k)&=& \frac{\pi\, \csc \left(\frac{2\, \pi\,  b}{\s^2}\right)\, \text{e}^{\frac{\pi\, a\, b}{2\, \sqrt{2~k}\, \s^3}} \Re\left[\Gamma \left(\frac{b\, \left(\frac{i\, \sqrt{2}\, a}{\sqrt{k}\, \s} - 2\right)}{2\, \s^2} + 1\right)\right]}{\Gamma \left(\frac{2\, b}{\s^2} - 1\right)\, \Gamma \left(\frac{b\, \left(\frac{i\, \sqrt{2}\, a}{\sqrt{k}\, \s} - 2\right)}{2\, \s^2} + 1\right)\, \Gamma \left(1 - \frac{b\, \left(\frac{i\, \sqrt{2}\, a}{\sqrt{k}\, \s} + 2\right)}{2\, \s^2}\right)}\, ,	\nonumber\\[1em]
\mathcal B(k) &= &-\frac{\text{e}^{-\frac{\pi\, a\, b}{2\, \sqrt{2~k}\, \s^3}}\,  \Re\left[\Gamma \left(\frac{b\, \left(2 - \frac{i\, \sqrt{2}\, a}{\sqrt{k}\, \s}\right)}{2\, \s^2}\right)\right]}{\Gamma \left(\frac{2\, b}{\s^2} - 1\right)} - \frac{\pi\, \csc \left(\frac{2\, \pi\, b}{\s^2}\right)\, \text{e}^{\frac{\pi\, a\, b}{2\, \sqrt{2~k}\, \s^3}}\, \Re\left[\Gamma \left(1 - \frac{b\, \left(\frac{i\, \sqrt{2}\, a}{\sqrt{k}\, \s} + 2\right)}{2\, \s^2}\right)\right]}{\Gamma\, \left(\frac{2\, b}{\s^2} - 1\right)\, \Gamma \left(\frac{b\, \left(\frac{i\, \sqrt{2}\, a}{\sqrt{k}\, \s} - 2\right)}{2\, \s^2} + 1\right)\, \Gamma \left(1 - \frac{b\, \left(\frac{i\, \sqrt{2}\, a}{\sqrt{k}\, \s} + 2\right)}{2\, \s^2}\right)}\, ,\nonumber \\[1em]
\mathcal C(k) &=& - \frac{\pi\, \csc \left(\frac{2\, \pi\,  b}{\s^2}\right)\, \text{e}^{\frac{\pi\, a\, b}{2\, \sqrt{2~k}\, \s^3}}\, \Im\left[\Gamma \left(\frac{b\, \left(\frac{i\, \sqrt{2}\, a}{\sqrt{k}\, \s} - 2\right)}{2\, \s^2} + 1\right)\right]}{\Gamma \left(\frac{2\, b}{\s^2} - 1\right)\, \Gamma \left(\frac{b\, \left(\frac{i\, \sqrt{2}\, a}{\sqrt{k}\, \s} - 2\right)}{2\, \s^2} + 1\right)\, \Gamma \left(1 - \frac{b\, \left(\frac{i\, \sqrt{2}\, a}{\sqrt{k}\, \s} + 2\right)}{2\, \s^2}\right)}\, , \nonumber\\[1em]
\mathcal D(k) &=& -\frac{\text{e}^{-\frac{\pi\,  a\, b}{2\, \sqrt{2~k}\, \s^3}}\, \Im\left[\Gamma \left(\frac{b\, \left(2 - \frac{i\, \sqrt{2}\, a}{\sqrt{k}\, \s}\right)}{2\,\s^2}\right)\right]}{\Gamma \left(\frac{2\, b}{\s^2}-1\right)}-\frac{\pi \, \csc \left(\frac{2\, \pi\,  b}{\s^2}\right)\, \text{e}^{\frac{\pi\, a\, b}{2\, \sqrt{2~k}\, \s^3}}\, \Im\left[\Gamma \left(1 - \frac{b\, \left(\frac{i\, \sqrt{2}\, a}{\sqrt{k}\, \s} + 2\right)}{2\, \s^2}\right)\right]}{\Gamma\left(\frac{2\, b}{\s^2} - 1\right)\, \Gamma \left(\frac{b\, \left(\frac{i\, \sqrt{2}\, a}{\sqrt{k}\, \s}-2\right)}{2\, \s^2} + 1\right)\, \Gamma \left(1 - \frac{b\, \left(\frac{i\, \sqrt{2}\, a}{\sqrt{k}\, \s} + 2\right)}{2\, \s^2}\right)}\, , \nonumber\\[1em]
\ea
in which $\Gamma$ is the Euler Gamma function, $\csc$ the cosecant, and $\Im$ and $\Re$ are the imaginary and real parts respectively.  We can sum the four terms as
\be
\{ \mathcal A(k)\, \text{e}^{\frac{i\, \pi\, b}{\s^2}} + \mathcal B(k)\, \text{e}^{i\, 0} + \mathcal C(k)\, \text{e}^{i\, \left(\frac{\pi\,  b}{\s^2} - \frac{\pi}{2}\right)} + \mathcal D(k)\, \text{e}^{- i\, \frac{\pi}{2}}\} = \mathcal Y(k)\, \text{e}^{i\, \mathcal W(k)}\, ,
\ee
where
\ba
\mathcal Y(k) = \left(\Re\, \left[\mathcal A(k)\, \text{e}^{\frac{i\, \pi\, b}{\s^2}} + \mathcal C(k)\, \text{e}^{i\, \left(\frac{\pi\, b}{\s^2} - \frac{\pi}{2}\right)} + \mathcal B(k)\, \text{e}^{i 0} + \mathcal D(k)\, \text{e}^{- i \frac{\pi}{2}}\right]^2 + \right. \nonumber \\ \left. 
\Im \left[\mathcal A(k)\, \text{e}^{\frac{i\, \pi\, b}{\s^2}} + \mathcal C(k)\, \text{e}^{i\, \left(\frac{\pi\, b}{\s^2} - \frac{\pi}{2}\right)} + \mathcal B(k)\, \text{e}^{i 0} + \mathcal D(k)\, \text{e}^{- i \frac{\pi}{2}}\right]^2 \right)^\fr \, , \nonumber
\ea
and
\ba
\mathcal W(k) = \arctan_2\left(\frac{\Im [\mathcal A(k)\, \text{e}^{\frac{i\, \pi\,  b}{\s^2}} + \mathcal B(k)\, \text{e}^{i\, 0} + \mathcal C(k)\, \text{e}^{i\, \left(\frac{\pi\,  b}{\s^2} - \frac{\pi}{2}\right)} + \mathcal D(k)\, \text{e}^{- i\, \frac{\pi}{2}}]}{\Re\, [\mathcal A(k)\, \text{e}^{\frac{i\, \pi\,  b}{\s^2}} + \mathcal B(k)\, \text{e}^{i\, 0} + \mathcal C(k)\, \text{e}^{i\, \left(\frac{\pi\,  b}{\s^2} - \frac{\pi}{2}\right)} + \mathcal D(k)\, \text{e}^{- i\, \frac{\pi}{2}}]}\right) \, ,\nonumber \\
\ea
in which $\text{arctan}_2$ is the two-argument arctangent. Therefore, the asymptotic behavior of the solution is 
\be
\lim_{x {\to} \infty}~\Re\, [K] = \mathcal Y(k)\, \cos(\phi_1(k,x) + \phi_2 (k,x) + \phi_3(k) + \mathcal W(k)) \, 
\ee
and the coefficient in Eq. \eqref{normhalf1} is the normalization  only if $\phi_1$ varies with $k$ more rapidly than does the amplitude $\mathcal Y$ and the other phases. As shown in Fig. \ref{fig:PlotPhases}, the approximation is robust, save for the small region near $k=0$, that contributes negligibly to the full integral over $k$. 
\end{appendices}
%

\end{document}